\documentclass{lmcs}
\pdfoutput=1

\usepackage{lastpage}
\lmcsdoi{16}{2}{10}
\lmcsheading{}{\pageref{LastPage}}{}{}%
{Nov.~30,~2018}{Jun.~03,~2020}{}

\usepackage[utf8]{inputenc}

\usepackage{graphicx}
\usepackage{booktabs}
\usepackage{graphics}
\usepackage{latexsym}
\usepackage{amssymb}
\usepackage{amsmath}
\usepackage{ifthen}
\usepackage{array}
\usepackage{mathrsfs}
\usepackage{tikz}
\usetikzlibrary{arrows, automata, positioning}
\renewcommand{\figurename}{Figure}
\usepackage{pgf, verbatim,eurosym}
\usetikzlibrary{arrows,automata}
\usepackage{sansmath}
\usepackage{dsfont}

\def\abs#1{\ensuremath{\lvert #1\rvert}}
\newcommand{\set}[1]{\{ #1 \}}
\newcommand{\halpha}{\widehat{\alpha}}
\newcommand{\A}{\mathcal{A}}
\newcommand{\C}{\mathcal{C}}
\newcommand{\D}{\mathcal{D}}
\newcommand{\E}{\mathcal{E}}
\newcommand{\R}{\mathcal{R}}

\newcommand{\Cond}{\mathit{Cond}}
\newcommand{\deltaD}{\delta_{\D}}
\newcommand{\deltaE}{\delta_{\E}}
\newcommand{\dists}{{\sf Dist}}
\newcommand{\eqmoves}{{\sf eqmoves}}
\newcommand{\Label}{{\sf L}}
\newcommand{\last}{{\sf last}}
\newcommand{\M}{\mathcal{M}}
\newcommand{\m}{{\sf m}}
\newcommand{\moves}{{\sf moves}}
\newcommand{\mem}{{\sf mem}}
\newcommand{\nat}{\mathbb{N}}
\newcommand{\norm}[1]{\lVert#1\rVert}
\newcommand{\Path}{{\sf Paths}}
\newcommand{\ExtPath}{{\sf ExtPaths}}
\newcommand{\post}{{\sf post}}
\newcommand{\Prob}{{\sf Pr}}
\newcommand{\Reals}{\mathbb{R}}
\newcommand{\sqm}{\sqsubseteq_{\sf m}}
\newcommand{\sqpm}{\sqsubseteq_{\sf pm}}
\newcommand{\subdis}{\mathit{subDist}}
\newcommand{\subdists}{{\sf subDist}}
\newcommand{\Succ}{{\sf Succ}}
\newcommand{\Supp}{{\sf Supp}}
\newcommand{\Tr}{\mathit{Tr}}
\newcommand{\trace}{{\sf trace}}
\newcommand{\tuple}[1]{\langle#1\rangle}
\newcommand{\V}{\mathcal{V}}

\newcommand{\fin}{{\mathit{fi}}}
\newcommand{\inn}{{\mathit{in}}}
\RequirePackage{tikz}
\usetikzlibrary{arrows}

\def\blue#1{\textcolor{blue}{#1}}

\renewcommand{\vec}[1]{\boldsymbol{#1}}
\begin{document}

\title[Trace Refinement in Labelled Markov Decision Processes]{\texorpdfstring{\vspace*{2mm}}{}Trace Refinement in \texorpdfstring{\\}{} Labelled Markov Decision Processes}\thanks{A preliminary version of this article appeared in the \emph{Proceedings of the 19th International Conference
on Foundations of Software Science and Computation Structures} (FoSSaCS), Lecture Notes in Computer Science 9634, Springer, 2016.}

\author[N.~Fijalkow]{Nathana\"{e}l Fijalkow\rsuper{{a,b}}}
\author[S.~Kiefer]{Stefan Kiefer\rsuper{c}}
\author[M.~Shirmohammadi]{Mahsa Shirmohammadi\rsuper{{c,d}}}

\address{\lsuper{a}CNRS, LaBRI, Bordeaux}
\address{\lsuper{b}Alan Turing Institute of data science, London}
\address{\lsuper{c}University of Oxford, UK}
\address{\lsuper{d}CNRS, IRIF, Paris}

\thanks{This work was supported by The Alan Turing Institute under the EPSRC grant EP/N510129/1 and the DeLTA project (ANR-16-CE40-0007).} 
\thanks{Stefan Kiefer is supported by the Royal Society.}

\begin{abstract}
Given two labelled Markov decision processes (MDPs), the trace-refinement problem
asks whether for all strategies of the first MDP there exists a strategy of the
second MDP such that the induced labelled Markov chains are trace-equivalent.
We show that this problem is decidable in polynomial time if the second MDP is a Markov chain.
The algorithm is based on new results on a particular notion of bisimulation
between distributions over the states.
However, we show that the general trace-refinement problem is undecidable, even if the first MDP is a Markov chain.
Decidability of those problems was stated as open in 2008.
We further study the decidability and complexity of the trace-refinement problem provided that
the strategies are restricted to be memoryless.
\end{abstract}

\maketitle

\section{Introduction}%
\label{sec-intro}
We consider labelled Markov chains (MCs)
whose transitions are labelled with symbols from an alphabet~$\Label$.
Upon taking a transition, the MC emits the associated label.
In this way, an MC defines a \emph{trace-probability} function $\Tr : \Label^* \to [0,1]$
which assigns to each finite trace $w \in \Label^*$ the probability that the MC emits~$w$ during its first $|w|$ transitions.
Consider the MC depicted in Figure~\ref{fig:undecidability_m1} with  initial  state~$p_0$.
For example, in  state~$p_0$,  with probability~$\frac{1}{4}$,
a transition to state~$p_c$ is taken and $c$ is emitted.
We have, e.g., $\Tr(a b c) = \frac{1}{4}\cdot \frac{1}{4}\cdot\frac{1}{4}$.
Two MCs over the same alphabet~$\Label$ are called \emph{equivalent} if their trace-probability functions are equal.

\begin{figure}
\centering
\renewcommand{\arraystretch}{1.3}
\centering
\begin{tikzpicture}[xscale=.8,>=latex',shorten >=1pt,node distance=3cm,on grid,auto]

 \node[state,initial,initial where=above,initial text={}] (p0) at (1,2) {$p_0$};
 \node[state] (pc) at (0,0) {$p_c$};
 \node[state] (pd) at (2,0) {$p_d$};
 \path[->] (p0) edge node [midway, left]  {$\frac{1}{4}, c$} (pc);
 \path[->] (p0) edge node [midway, right]  {$\frac{1}{4}, d$} (pd);

\path[->] (p0) edge  [out=165,in=195,looseness=8] node [midway, left] {$\frac{1}{4}, a$} (p0);
 \path[->] (p0) edge [out=15,in=-15,looseness=8]  node [midway, right]{$\frac{1}{4}, b$} (p0);

 \path[->] (pc) edge  [out=165,in=195,looseness=8] node [midway, left] {$1, c$} (pc);
 \path[->] (pd) edge [out=15,in=-15,looseness=8]  node [midway, right]{$1, d$} (pd);

 \node[label]  at (10,1.2) {\begin{tabular}{ l c }
  trace~$w$ & $\Tr(w)$ \\
	\toprule
  $w\in (a+b)^{n}$ & $\frac{1}{4^{n}}$\\
  $w\in (a+b)^{n}c^{+}$ & $\frac{1}{4^{n+1}}$ \\
	$w\in (a+b)^{n}d^{+}$ & $\frac{1}{4^{n+1}}$ \\
	\text{otherwise} & 0\\
\end{tabular}};
\end{tikzpicture}

\caption{An MC with its  trace-probability function.
This MC, denoted by~$\C(\A)$, will also be used in Section~\ref{sec-undecidability-results} for the reduction from universality of probabilistic automata to the trace-refinement problem.}%
\label{fig:undecidability_m1}
\end{figure}

The study of labelled MCs and their equivalence has a long history, going back to Sch\"utzenberger~\cite{Schutzenberger} and Paz~\cite{Paz71}
who studied \emph{weighted} and \emph{probabilistic} automata, respectively.
Those models generalize labelled MCs, but the respective equivalence problems are essentially the same.
It can be extracted from~\cite{Schutzenberger} that equivalence is decidable in polynomial time, using a technique based on linear algebra.
Variants of this technique were developed in~\cite{Tzeng,DoyenHR08}.
Tzeng~\cite{Tzeng96} considered the path-equivalence problem for nondeterministic
automata which asks, given nondeterministic automata $A$~and~$B$, whether each
word has the same number of accepting paths in~$A$ as in~$B$. He gives an {\sf NC}
algorithm\footnote{%
The complexity class~{\sf NC} is the subclass of~{\sf P} containing those problems that can be solved in polylogarithmic parallel time (see, e.g.,~\cite{GHR95}).}
for deciding path equivalence which can be straightforwardly adapted to yield
an {\sf NC} algorithm for equivalence of MCs.

More recently, the efficient decidability of the equivalence problem was exploited, both theoretically and practically, for the verification of probabilistic systems, see, e.g.,~\cite{KMOWW:CAV11,12KMOWW:CAV,Peyronnet12,Ngo13,Li15}.
In those works, equivalence naturally expresses properties such as obliviousness and anonymity, which are difficult to formalize
in temporal logic.
The \emph{inclusion problem} for two probabilistic automata asks whether for each word the acceptance probability in the first automaton is less than or equal to the acceptance probability in the second automaton.
Despite its semblance to the equivalence problem,
the inclusion problem is undecidable~\cite{CL89}, even for automata of fixed dimension~\cite{BC03}.
This is unfortunate, especially because deciding language inclusion is often at the heart of verification algorithms.

We study another ``inclusion-like'' generalization of the equivalence problem: trace refinement in labelled Markov decision processes (MDPs).
MDPs extend MCs by nondeterminism,
and labelled MDPs generate outputs (labels);
thus labelled MDPs are a generative model with nondeterminism.
In each state, a controller chooses, possibly randomly and possibly depending on the history, one out of finitely many \emph{moves}%
\footnote{As in~\cite{DoyenHR08} we speak of moves rather than of actions,
to avoid possible confusion with the  label alphabet~$\Label$.}%
.
A move determines a probability distribution over the emitted label and the successor state.
In this way, an MDP and a strategy of the controller induce an MC\@.

The \emph{trace-refinement problem} asks,  given two MDPs $\D$ and $\E$,
whether for all strategies for~$\D$ there is a strategy for~$\E$ such that the induced MCs are equivalent.
Consider the MDP depicted in Figure~\ref{fig:introMDP} where in state~$q_1$ there are two available moves;
 one move generates the label~$c$ with probability 1, the other move generates~$d$ with probability 1.
A strategy of the controller that, in  state~$q_1$, chooses the last generated label (either $c$ or~$d$) with probability~$1$,
induces the same trace-probability function as the MC shown in Figure~\ref{fig:undecidability_m1}; the MDP
thus refines that MC\@.
The described strategy  needs one bit of memory to keep track of the last generated label.
It was  shown in~\cite{DoyenHR08} that the strategy for~$\E$  may require \emph{infinite memory}, even if $\D$ is an MC\@.
The decidability of trace refinement was posed as an open problem, both in the introduction and in the conclusion of~\cite{DoyenHR08}.
The authors of~\cite{DoyenHR08} also ask about the decidability of subcases, where $\D$ or~$\E$ are restricted to be MCs.
In this paper we answer all those questions.
We show that trace refinement is undecidable, even if $\D$ is an MC\@.
In contrast, we show that trace refinement is decidable efficiently (in~{\sf NC}, hence in~{\sf P}), if $\E$ is an MC\@.
Moreover, we prove that the trace-refinement problem becomes decidable if one imposes suitable \emph{restrictions on the strategies} for $\D$ and~$\E$, respectively.
More specifically, we consider \emph{memoryless} (i.e., no dependence on the history) and \emph{pure memoryless} (i.e., no randomization and no dependence on the history)  strategies,
establishing various complexity results between {\sf NP} and~{\sf PSPACE}.

\begin{figure}
\centering
\tikzstyle{BoxStyle} = [draw, circle, fill=black, scale=0.4,minimum width = 1pt, minimum height = 1pt]

\centering
\begin{tikzpicture}[xscale=.8,>=latex',shorten >=1pt,node distance=3cm,on grid,auto]

 \node[state,initial,initial where=above,initial text={}] (q0) at (0,2) {$q_0$};
 \node[state] (q1) at (0,0) {$q_1$};
 \path[->] (q0) edge [bend left=-20] node [pos=0.4, left]  {$\frac{1}{4}, c$} (q1);
 \path[->] (q0) edge [bend left=20] node [pos=0.4, right]  {$\frac{1}{4}, d$} (q1);

\path[->] (q0) edge  [out=165,in=195,looseness=8] node [midway, left] {$\frac{1}{4}, a$} (q0);
 \path[->] (q0) edge [out=15,in=-15,looseness=8]  node [midway, right]{$\frac{1}{4}, b$} (q0);

 \path[->] (q1) edge  [out=165,in=195,looseness=8] node [midway, left] {$1, c$} (q1);
 \node[BoxStyle,label={{$\m_{1}$}}] at (-0.9,0.15){};
 \node[BoxStyle,label={{$\m_{2}$}}] at (0.9,0.15){};

 \path[->] (q1) edge [out=15,in=-15,looseness=8]  node [midway, right]{$1, d$} (q1);
\end{tikzpicture}

\caption{An MDP where the  choice of controller is  relevant only in~$q_1$.
Two available moves~$\m_1,\m_2$ are shown with small black circles.}%
\label{fig:introMDP}
\end{figure}

To obtain the aforementioned {\sf NC} result, we demonstrate a link between trace refinement
and a particular notion of \emph{bisimulation} between two MDPs that was studied in~\cite{Jans}.
This variant of bisimulation is not defined between two states as in the usual notion, but between two \emph{distributions} on states.
An exponential-time algorithm that decides (this notion of) bisimulation was provided in~\cite{Jans}.
We sharpen this result by exhibiting
a {\sf coNP} algorithm that decides bisimulation between two MDPs,
and an {\sf NC} algorithm for the case where one of the MDPs is an MC\@.
For that we refine the arguments devised in~\cite{Jans}.
The model considered in~\cite{Jans} is more general than ours in that they also consider continuous state spaces, but more restricted than ours in that the label is determined by the move.


\section{Preliminaries}%
\label{sec-preliminaries}
A {trace} over a finite set~$\Label$ of labels is a finite sequence~$w=a_1 \cdots a_n$
of labels where the length of the trace is~$\abs{w}=n$.
The empty trace~$\epsilon$ has length zero.
For~$n\geq 0$, let $\Label^{n}$ be the set of all traces with length~$n$;
we denote by~$\Label^{*}$ the set of all (finite) traces over~$\Label$.

For a function~$d : S \to [0, 1]$ over a countable set~$S$,
define the \emph{norm} $\norm{d} := \sum_{s \in S} d(s)$.
The \emph{support} of~$d$ is the set $\Supp(d) = \{s \in S \mid d(s) > 0\}$.
The function~$d$ is a \emph{probability subdistribution} over~$S$ if $\norm{d} \le 1$;
it is a \emph{probability distribution} if~$\norm{d} = 1$.
We denote by $\subdists(S)$ (resp.~$\dists(S)$) the set of all probability subdistributions (resp.\ distributions) over~$S$.
Given $s \in S$, the \emph{Dirac distribution} on~$s$ assigns probability~$1$
to~$s$;  we  denote it  by~$d_s$.
For a non-empty finite subset~$T \subseteq S$, the \emph{uniform distribution} over $T$ assigns probability~$\frac{1}{\abs{T}}$ to every element in $T$.

\subsection{Labelled Markov Decision Processes}

In this article a \emph{labelled Markov decision process} (MDP) is a generative probabilistic model with nondeterminism.
Formally, an MDP is a quadruple $\D = \tuple{Q,\mu_0,\Label,\delta}$,
consisting of a finite set~$Q$ of states,  an initial distribution~$\mu_0 \in \dists(Q)$,
a finite set~$\Label$ of labels,
and a finite probabilistic transition relation~$\delta \subseteq Q \times \dists(\Label \times Q)$ where
states are in relation with distributions over pairs of labels and successors.
We assume that for each state $q\in Q$ there exists some distribution
$d\in \dists(\Label \times Q)$ where~$\tuple{q,d} \in \delta$.
The set of \emph{moves} in~$q$ is~$\moves(q) = \{d\in \dists(\Label \times Q) \mid \tuple{q,d} \in \delta \}$;
denote by $\moves=\bigcup_{q\in Q} \moves(q)$  the set of all moves.

For the complexity results, we assume that  probabilities of transitions are rational and
given as fractions of integers represented in binary.

We describe the behaviour of an MDP as a trace generator running in steps.
The MDP starts in the first step in state $q$ with probability~$\mu_{0}(q)$.
In each step, if the MDP is in state $q$
the controller chooses~$\m\in\moves(q)$;
then, with probability~$\m(a,q')$, the label~$a$ is generated and the next step starts in the
successor state~$q'$.

Given $q\in Q$, denote by $\post(q)$ the set
$\{(a,q') \in  \Supp(\m) \mid \m \in \moves(q)\}$.
A \emph{path} in $\D$ is a sequence $\rho=q_{0} a_{1} q_{1} \dots a_n q_{n}$
such that  $(a_{i+1},q_{i+1})\in \post(q_{i})$ for all $0\leq i<n$.
The last state of $\rho$ is $\last(\rho) = q_n$.
The trace $\trace(\rho)$ generated by $\rho$ is~$a_1 a_2 \cdots a_n$.
We let $\Path(\D)$ denote the set of paths in~$\D$,
$\Path(w)$ denote $\{\rho\in \Path(\D) \mid \trace(\rho) = w\}$ the set of paths generating~$w$,
and $\Path(w,q)$ denote $\{\rho\in \Path(\D) \mid \trace(\rho) = w \text{ and } \last(\rho) = q\}$ the set of paths generating~$w$
ending in~$q$.

\paragraph{Strategies.}
A \textit{strategy} for an MDP $\D$  is a function
$\alpha: \Path(\D) \to \dists(\moves)$ that, given a path $\rho$,
returns a probability distribution $\alpha(\rho) \in \dists(\moves(\last(\rho)))$.
Let $q = \last(\rho)$, then $\alpha(\rho)$ generates a label~$a$ and selects a successor state $q'$
with probability
\[
\sum_{\m \in \moves(q)} \alpha(\rho)(\m) \cdot \m(a,q').
\]
Abusing notation slightly we write $\alpha(\rho)(a,q')$
for $\sum_{\m \in \moves} \alpha(\rho)(\m) \cdot \m(a,q')$.

A strategy~$\alpha$ is \emph{pure} if for all $\rho \in \Path(\D)$,
there exists~$\m \in \moves$ such that $\alpha(\rho)(\m)=1$;
we thus view pure strategies as functions~$\alpha: \Path(\D)\to \moves$.
A strategy~$\alpha$ is \emph{memoryless} if $\alpha(\rho) = \alpha(\rho')$
for all paths~$\rho, \rho'$ with $\last(\rho) = \last(\rho')$;
we thus view memoryless strategies as functions $\alpha: Q \to \dists(\moves)$.
A strategy~$\alpha$ is \emph{trace-based} if  $\alpha(\rho) = \alpha(\rho')$
for all~$\rho, \rho'$ where $\trace(\rho)=\trace(\rho')$ and $\last(\rho) = \last(\rho')$;
we  view trace-based strategies as functions~$\alpha:\Label^{*} \times Q \to \dists(\moves)$.
For a traced-based strategy~$\alpha$  we write $\alpha(w,q)(a,q')$
for $\sum_{\m \in \moves} \alpha(w,q)(\m) \cdot \m(a,q')$.

\paragraph{Trace-probability function.}
For an MDP~$\D$ and a strategy~$\alpha$,
the probability of a single path is
inductively defined by $\Prob_{\D,\alpha}(q) = \mu_0(q)$ and
\[
\Prob_{\D,\alpha}(\rho a q') = \Prob_{\D,\alpha}(\rho) \cdot \alpha(\rho)(a,q').
\]
This induces
$\Prob_{\D,\alpha}(w,q) = \sum_{\rho \in \Path(w,q)} \Prob_{\D,\alpha}(\rho)$
and $\Prob_{\D,\alpha}(w) = \sum_{\rho \in \Path(w)} \Prob_{\D,\alpha}(\rho)$.

The \emph{trace-probability} function $\Tr_{\D,\alpha} : \Label^{*} \to [0,1]$ is, given
a trace~$w$, defined by
\[\Tr_{\D,\alpha}( w ) = \Prob_{\D,\alpha}(w).\]
We may drop the subscript $\D$ or~$\alpha$ from~$\Tr_{\D,\alpha}$ and from $\Prob_{\D,\alpha}$ if it is understood.
We let $\subdis_{\D,\alpha}(w) \in \subdists(Q)$ denote the subdistribution after generating a trace~$w$, that is
\[
\subdis_{\D,\alpha}(w)(q)= \Prob_{\D,\alpha}(w,q).
\]

We have:
\begin{equation} \label{eq-subdis=trace-probability}
\Tr_{\D,\alpha}( w ) =\norm{\subdis_{\D,\alpha}(w)}
\end{equation}

\medskip
Let $\D$ be an MDP and $\alpha,\beta$ be two strategies; we denote by $\Tr_{\alpha} = \Tr_{\beta}$
when the equality
$\Tr_{\alpha}(w) = \Tr_{\beta}(w)$ holds for all traces~$w \in \Label^*$.
A version of the following lemma was proved in~\cite[Lemma~1]{DoyenHR08}:

\begin{lem}%
\label{lem-trace-based-enough}
Let $\D$ be an MDP and $\alpha$ be a strategy.
There exists a trace-based strategy~$\beta$ such that
$\Tr_{\alpha} = \Tr_{\beta}$.
\end{lem}
\begin{proof}
Let $\alpha$ be a strategy $\alpha : \Path \to \dists(\moves)$ of the MDP~$\D = \tuple{Q,\mu_0,\Label,\delta}$.
We define a trace-based strategy~$\beta: \Label^{*} \times Q \to \dists(\moves)$ as follows:
given a pair of state~$q$ and trace~$w$, let
\[
\beta(w,q) = \frac{\sum_{\rho \in \Path(w,q)} \Prob_{\alpha}(\rho) \cdot \alpha(\rho)}{\Prob_{\alpha}(w,q)}\,.
\]
We prove by induction that $\Prob_{\beta}(w,q) = \Prob_{\alpha}(w,q)$.
The induction base, $w = \epsilon$\blue{,} is simple.

For any $w, q'$ we have:
\begin{align*}
\Prob_{\beta}(wa,q') & = \sum_{q \in Q} \Prob_{\beta}(w,q) \cdot \beta(w,q)(a,q') \\
 & = \sum_{q \in Q} \Prob_{\alpha}(w,q) \cdot \frac{\sum_{\rho \in \Path(w,q)} \Prob_{\alpha}(\rho) \cdot \alpha(\rho)(a,q')}{\Prob_{\alpha}(w,q)} \\
 & = \sum_{q \in Q} \sum_{\rho \in \Path(w,q)} \Prob_{\alpha}(\rho) \cdot \alpha(\rho)(a,q') \\
 & = \sum_{\rho \in \Path(w)} \Prob_{\alpha}(\rho) \cdot \alpha(\rho)(a,q') \\
 & = \Prob_{\alpha}(wa,q'). \qedhere
\end{align*}
\end{proof}


A strategy can be implemented by means of memory; to
make this explicit, we define  a variant of the notion of strategies.
We instrument strategies  with a countable set~$\mem$ of \emph{memory modes}.
For an MDP~$\D$, a \emph{generalized strategy} with memory $\mem$ is defined by
$d_0 \in \dists(\mem)$ and $\alpha: \mem \times \Path(\D) \to \dists(\moves \times \mem)$.
The strategy~$\alpha$ returns a probability distribution over
the next moves and next memory modes based on the  taken path to the current state
 and the current memory mode.
We show in Lemma~\ref{lem-non-general-enough} that for each generalized strategy~$\alpha$ there is
a strategy~$\beta$ such that
each path in the MDP is equally probable under
both strategies~$\alpha$ and $\beta$, implying that the trace-probability functions induced by~$\alpha$ and $\beta$ are also equal.

To formalize generalized strategies, we extend the definitions:
An extended path is a sequence $\pi = q_0 M_0 a_1 q_1 M_1 \dots a_n q_n M_n$
such that  $(a_{i+1},q_{i+1})\in \post(q_{i})$ for all $0\leq i<n$.
The last memory state is $\last_\mem(\pi) = M_n$.
All notions for paths are naturally transferred to extended paths.
For an extended path~$\pi=q_0 M_0 a_1 q_1 M_1 \dots a_n q_n M_n$, let
$\rho_{\pi}$ denote
its projection to a path, that is $\rho_{\pi}=q_0  a_1 q_1  \dots a_n q_n$.
We let $\ExtPath$ be the set of extended paths, $\ExtPath(\rho)$ be the set of extended paths projecting to~$\rho$,
and $\ExtPath(\rho,M)$ be the set of extended paths~$\pi$ projecting to~$\rho$
and such that $\last_\mem(\pi) = M$.

Given a memory mode~$M$ and a path~$\rho\in \Path(\D)$,
we write $\alpha(M,\rho)(a,q',M')$
for $\sum_{\m \in \moves} \alpha(M,\rho)(\m,M') \cdot \m(a,q')$.
The probability of an extended path $\pi$ is defined inductively by
$\Prob_{\D,\alpha}(qm) = \mu_0(q) \cdot d_0(m)$   and
\[
\Prob_{\D,\alpha}(\pi a q' M') = \Prob_{\D,\alpha}(\pi)
\cdot \alpha(\last_\mem(\pi),\rho_{\pi})(a,q',M').
\]
We then let
\[\Prob_{\D,\alpha}(\rho) = \sum_{\pi \in \ExtPath(\rho)} \Prob_{\D,\alpha}(\pi) \quad \text{ and } \quad
\Prob_{\D,\alpha}(\rho,M) = \sum_{\pi \in \ExtPath(\rho,M)} \Prob_{\D,\alpha}(\pi).\]

\color{black}

It easily follows from these definitions that
\[
\Prob_{\D,\alpha}(\rho a q') = \sum_{M \in \mem} \Prob_{\D,\alpha}(\rho,M)
 \cdot \alpha(M,\rho)(a,q').
\]

\begin{lem}%
\label{lem-non-general-enough}
Let $\D$ be an MDP and $\alpha$ be a generalized strategy.
There exists a strategy~$\beta$ such that
$\Tr_{\alpha} = \Tr_{\beta}$.
\end{lem}

\begin{proof}
Let $\D = \tuple{Q,\mu_0,\Label,\delta}$ be an MDP and
 $\alpha: \mem \times \Path(\D) \to \dists(\moves \times \mem)$ be a
generalized strategy of~$\D$.
We drop the subscript~$\D$ in the rest of the proof.

Define a strategy~$\beta: \Path \to \dists(\moves)$  from $\alpha$ as follows. Given a path~$\rho$, let
\[
\beta(\rho) = \frac{\sum_{M\in \mem} \Prob_{\alpha}(\rho,M) \cdot \alpha(M,\rho)}{\Prob_{\alpha}(\rho)}\,.
\]
We prove by induction that $\Prob_{\beta}(\rho) = \Prob_{\alpha}(\rho)$.
The induction base, $\rho = q$ for  $q \in Q$, is simple.
For all $\rho, a, q'$ we have:
\begin{align*}
\Prob_{\beta}(\rho a q') &= \Prob_{\beta}(\rho) \cdot \beta(\rho)(a,q') \\
 & = \Prob_{\alpha}(\rho) \cdot \frac{\sum_{M \in \mem} \Prob_{\alpha}(\rho,M) \cdot \alpha(M,\rho)(a,q')}{\Prob_{\alpha}(\rho)} \\
 & = \sum_{M\in \mem} \Prob_{\alpha}(\rho,M) \cdot \alpha(M,\rho)(a,q') \\
 & = \Prob_{\alpha}(\rho a q').
\end{align*}

Since the probability of a trace~$w$ is a summation over the probability of all paths emitting~$w$,
having $\Prob_{\alpha}(\rho) = \Prob_{\beta}(\rho)$ for all $\rho$ implies that $\Tr_{\alpha} = \Tr_{\beta}$.
\end{proof}

\paragraph{Labelled Markov Chains.}
A finite-state labelled Markov chain (MC for short) is an MDP where
only a single move is available in each state, and thus controller's choice plays no role.
An MC  $\C = \tuple{Q,\mu_0,\Label,\delta}$ is an MDP where~$\delta: Q \to \dists(\Label \times Q)$
is a  probabilistic transition function.
Since MCs are MDPs, we analogously define paths, and the probability of a single path
inductively as follows:
$\Prob_{\C}(q) = \mu_0(q)$ and $\Prob_{\C}(\rho a q) = \Prob_{\C}(\rho) \cdot \delta(q')(a,q)$
where $q'=\last(\rho)$.
The notations~$\subdis_{\C}(w)$ and~$\Tr_{\C}$ are defined analogously.


\subsection{Trace Refinement}%
\label{sub-prel-trace-refine}

Given two MDPs~$\D$ and $\E$ with the same set~$\Label$ of labels, we say that
\emph{$\E$ refines~$\D$}, denoted by $\D \sqsubseteq \E$,
if  for all strategies~$\alpha$ for~$\D$ there exists some strategy~$\beta$ for~$\E$ such that
$\Tr_{\D} = \Tr_{\E}$.
We are interested in the problem ${\sf MDP\sqsubseteq MDP}$, which asks, for two given MDPs~$\D$ and~$\E$, whether $\D \sqsubseteq \E$.
The decidability of this problem was posed as an open question in~\cite{DoyenHR08}.
We show in Theorem~\ref{theo:unde-MDP-MDP} that the problem~${\sf MDP\sqsubseteq MDP}$ is undecidable.

We consider various subproblems of ${\sf MDP\sqsubseteq MDP}$, which asks whether $\D \sqsubseteq \E$ holds.
Specifically, we speak of the problem
\begin{itemize}
  \item ${\sf MDP\sqsubseteq MC}$ when $\E$ is restricted to be an MC;\@
  \item ${\sf MC\sqsubseteq MDP}$ when $\D$ is restricted to be an MC;\@
  \item ${\sf MC\sqsubseteq MC}$ when both $\D$~and~$\E$ are restricted to be MCs.
\end{itemize}
We show in Theorem~\ref{theo:unde-MDP-MDP} that even the problem ${\sf MC\sqsubseteq MDP}$ is undecidable.
Hence we consider further subproblems.
Specifically, we denote by~${\sf MC\sqm MDP_{}}$ the problem where the MDP is restricted to use only memoryless strategies,
and by~${\sf MC\sqpm MDP_{}}$ the problem where the MDP is restricted to use only pure memoryless strategies.
When both MDPs~$\D$ and~$\E$ are restricted to use only pure memoryless strategies, the trace-refinement problem is denoted by~${\sf MDP_{pm}\sqpm MDP_{pm}}$.
The problem ${\sf MC\sqsubseteq MC}$ equals the \emph{trace-equivalence problem} for MCs: given two   MCs $\C_1,\C_2$ we have $\C_1 \sqsubseteq C_2$ if and only if $\Tr_{\C_1} = \Tr_{\C_2}$ if and only if $\C_2 \sqsubseteq \C_1$.
This problem is known to be in {\sf NC}~\cite{Tzeng96}, hence in~{\sf P}.


\section{Undecidability Results}%
\label{sec-undecidability-results}
In this section we show:
\begin{thm}\label{theo:unde-MDP-MDP}
The problem~${\sf MC\sqsubseteq MDP}$ is undecidable.
Hence a fortiori, ${\sf MDP\sqsubseteq MDP}$ is undecidable.
\end{thm}

\begin{proof}
\newcommand{\dis}{\mathit{dis}}%
To show that the problem~${\sf MC \sqsubseteq MDP}$ is undecidable,
we  establish a reduction from the universality problem for probabilistic automata.
A \emph{probabilistic automaton} is a tuple $\A = \tuple{Q,\mu_0,\Label,\delta,F}$
consisting of a finite set $Q$ of states, an initial distribution $\mu_0 \in \dists(Q)$,
a finite set $\Label$ of letters,
a transition function $\delta : Q \times \Label \to \dists(Q)$ assigning to every state and letter a distribution over states,
and a set~$F$ of final states.
For a word $w \in \Label^*$ we write $\dis_\A(w) \in \dists(Q)$ for the distribution such that, for all $q \in Q$, we have that $\dis_\A(w)(q)$ is the probability that after inputting~$w$ the automaton~$\A$ is in state~$q$.
We write $\Prob_\A(w) = \sum_{q \in F} \dis_\A(w)(q)$ to denote the probability that $\A$ accepts~$w$.
The \emph{universality problem} asks, given a probabilistic automaton $\A$, whether $\Prob_\A(w) \ge \frac{1}{2}$ holds for all words~$w$.
This problem is known to be undecidable~\cite{Paz71,Fijalkow17}.

Let $\A = \tuple{Q,\mu_0,\Label,\delta,F}$ be a probabilistic automaton; without loss of generality we assume that $\Label = \set{a,b}$.
We construct an MDP~$\D$ such that $\A$ is universal  if and only if $\C \sqsubseteq \D$ where
$\C$ is the MC shown in Figure~\ref{fig:undecidability_m1}.

The MDP~$\D$ is constructed from $\A$ as follows; see Figure~\ref{fig:undecidabilitySketch}.
Its set of states is $Q \cup \set{q_c,q_d}$, and its initial distribution is~$\mu_0$.
(Here and in the following we identify subdistributions $\mu \in \subdists(Q)$ and $\mu \in \subdists(Q \cup \{q_c,q_d\})$ if $\mu(q_c) = \mu(q_d) = 0$.)
We describe the transitions of~$\D$ using the transition function~$\delta$ of~$\A$.
Consider a state $q \in Q$:
\begin{itemize}
	\item If $q \in F$, there are two available moves~$\m_c,\m_d$; both emit $a$ with probability $\frac{1}{4}$ and simulate the probabilistic automaton $\A$
	reading 	the letter $a$, 	or emit $b$ with probability $\frac{1}{4}$ and simulate the probabilistic automaton $\A$
	reading 	the letter $b$.
	With the remaining probability of $\frac{1}{2}$, $\m_c$ emits $c$ and leads to $q_c$ and $\m_d$ emits $d$ and leads to $q_d$.
	Formally, $\m_c(c,q_c)=\frac{1}{2}$, $\m_d(d,q_d)=\frac{1}{2}$ and $\m_c(e,q')=\m_d(e,q')=\frac{1}{4}\delta(q,e)(q')$ where $q' \in Q$ and $e\in \{a,b\}$.
	\item If~$q \notin F$,  there is a single available move~$\m$  such that
	$\m(d,q_d)=\frac{1}{2}$ and $\m(e,q')=\frac{1}{4}\delta(q,e)(q')$ where~$q'\in Q$ and $e\in \{a,b\}$.
	\item The only move from $q_c$ is the Dirac distribution on $(c,q_c)$; likewise the only move from $q_d$ is the Dirac distribution on $(d,q_d)$.
\end{itemize}
This MDP~$\D$ ``is almost'' an MC, in the sense that a strategy~$\alpha$ does not influence its behaviour until eventually a transition to $q_c$ or~$q_d$ is taken.
Indeed, for all~$\alpha$ and for all $w \in {\{a,b\}}^*$ we have $\subdis_{\D,\alpha}(w) = \frac{1}{4^{|w|}}\dis_\A(w)$.
In particular, it follows $\Tr_{\D,\alpha}(w) = \norm{\subdis_{\D,\alpha}(w)} = \frac{1}{4^{|w|}} \norm{\dis_\A(w)} = \frac{1}{4^{|w|}}$.
Further, if $\alpha$ is trace-based we have:
\begin{equation}
\begin{aligned}
 \Tr_{\D,\alpha}(w c)
 & = \norm{\subdis_{\D,\alpha}(w c)}
  && \text{by~\eqref{eq-subdis=trace-probability}} \\
 & = \subdis_{\D,\alpha}(w c)(q_c)
  && \text{structure of~$\D$} \\
 & = \sum_{q \in F} \subdis_{\D,\alpha}(w)(q) \cdot \alpha(w, q)(\m_c) \cdot \frac12
  && \text{structure of~$\D$} \\
 & = \frac{1}{4^{|w|}} \sum_{q \in F} \dis_\A(w)(q) \cdot \alpha(w, q)(\m_c) \cdot \frac12
  && \text{as argued above}
\end{aligned}
\label{eq-undec-wd}
\end{equation}

\begin{figure}[t]
\begin{center}
    \centering

\tikzstyle{BoxStyle} = [draw, circle, fill=black, scale=0.4,minimum width = 1pt, minimum height = 1pt]
\begin{tikzpicture}[xscale=.8,yscale=1.2,>=latex',shorten >=1pt,node distance=3cm,on grid,auto]

\draw (-1.3,-.2) [dashed] rectangle (3.3,4.4);

\node[label]  at (.8,4.7) {the automaton~$\A$};
\node[state] (q) at (-.4,1) {$q$};
\node[state] (qx) at (2.5,1.5) {$q_1$};
\node[state] (qy) at (2.5,0.3) {$q_2$};
\path[->] (q) edge node  [midway,above] {$x',a$} (qx);
\path[->] (q) edge node  [midway,below] {$y',b$} (qy);

\node[state, accepting] (p) at (-.4,3.3) {$p$};
\node[state] (px) at (2.5,3.9) {$p_1$};
\node[state] (py) at (2.5,2.7) {$p_2$};
\path[->] (p) edge node  [midway,above] {$x,a$} (px);
\path[->] (p) edge node  [midway,below] {$y,b$} (py);

\node[label]  at (3.8,2.2) {$\Longrightarrow$};

\draw (4.2,-.25) [densely dotted] rectangle (15.5,4.7);
\node[label]  at (5.7,5.0) {the MDP~$\D$};
\draw (4.7,-.2) [dashed] rectangle (10.1,4.4);

 \node[state] (c) at (13,2.7) {$q_d$};
\node[state] (d) at (13,3.9) {$q_c$};

\path[->] (c) edge [loop right] node [right, midway] {$1,d$}  (c);
\path[->] (d) edge [loop right] node [right, midway] {$1,c$}  (d);

\node[state] (q) at (5.6,1) {$q$};
\node[BoxStyle,label=above:{{$\m$}}] (m) at (6.7,1){};
\node[state] (qx) at (9.3,1.5) {$q_1$};
\node[state] (qy) at (9.3,0.3) {$q_2$};
\path[-] (q) edge (m.center);
\path[->] (m) edge node  [midway, above] {$\frac{x'}{4},a$} (qx);
\path[->] (m) edge node  [midway, below] {$\frac{y'}{4},b$} (qy);
\path[->] (m) edge [bend left=-20] node [near end, below,yshift=-1mm] {$\frac{1}{2},d$} (c);

\node[state] (p) at (5.6,3.3) {$p$};
\node[BoxStyle,label=above:{{$\m_c$}}] (m2) at (6.7,3.9){};
\path[-] (p) edge (m2.center);
\node[state] (px) at (9.3,3.9) {$p_1$};
\node[state] (py) at (9.3,2.7) {$p_2$};
\path[->] (m2) edge node  [midway, below= .4cm of m2] {$~\frac{x}{4},a$}(px);
\path[-] (m2.center) edge (7,2.76) ;
\path[->] (7,2.8) edge  (8.7,2.8);
\path[->] (m2) edge [bend left=20] node [near end, below] {$\frac{1}{2},c$} (d);
\node[BoxStyle,label=below:{{$\m_d$}}] (m1) at (6.7,2.7){};
\path[-] (p) edge (m1.center);
\path[-] (m1.center) edge (7,3.84) ;
\path[->] (7,3.8) edge (8.7,3.8);
\path[->] (m1) edge node   [midway, above= .4cm of m1] {$~\frac{y}{4},b$} (py);
\path[->] (m1) edge [bend left=-20] node [near end, above] {$\frac{1}{2},d$} (c);

\end{tikzpicture}

\end{center}
 \caption{
Sketch of the construction of the MDP~$\D$ from the probabilistic automaton~$\A$,
for the undecidability result of~${\sf MC \sqsubseteq MDP}$.
Here, $p$ is an accepting state whereas $q$ is not.
To read the picture, note that in~$p$ there is a transition to the state $p_1$ with probability~$x$ and label~$a$: $\delta(p,a)(p_1)=x$.
}\label{fig:undecidabilitySketch}
\end{figure}

We show that $\A$ is universal if and only if $\C \sqsubseteq \D$.
Let $\A$ be universal.
Define a trace-based strategy~$\alpha$ with $\alpha(w, q)(\m_c) = \frac{1}{2 \Prob_\A(w)}$ for all $w \in {\{a,b\}}^*$ and $q \in F$.
Note that $\alpha(w, q)(\m_c)$ is a probability as $\Prob_\A(w) \ge \frac12$.
Let $w \in {\{a, b\}}^*$.
We have:
\begin{align*}
\Tr_{\D, \alpha}(w)
 &= \frac{1}{4^{|w|}} && \text{as argued above} \\
 &= \Tr_{\C}(w) && \text{Figure~\ref{fig:undecidability_m1}}
\end{align*}
Further we have:
\begin{align*}
\Tr_{\D, \alpha}(w c)
 &= \frac{1}{4^{|w|}} \sum_{q \in F} \dis_\A(w)(q) \cdot \alpha(w, q)(\m_c) \cdot \frac12
  && \text{by~\eqref{eq-undec-wd}} \\
 &= \frac{1}{4^{|w|}} \sum_{q \in F} \dis_\A(w)(q) \cdot \frac{1}{\Prob_\A(w)} \cdot \frac14
  && \text{definition of~$\alpha$} \\
 &= \frac{1}{4^{|w|+1}}
  && \Prob_\A(w) = \sum_{q \in F} \dis_\A(w)(q) \\
 &= \Tr_{\C}(w c)  && \text{Figure~\ref{fig:undecidability_m1}}
\end{align*}
It follows from the definitions of $\D$ and~$\C$ that  for all $k \ge 1$, we have $\Tr_{\D,\alpha}(w c^k) = \Tr_{\D,\alpha}(w c) = \Tr_{\C}(w c) = \Tr_{\C}(w c^k)$.
We have $\sum_{e \in \{a,b,c,d\}} \Tr_{\D,\alpha}(w e) = \Tr_{\D,\alpha}(w) = \Tr_{\C}(w) = \sum_{e \in \{a,b,c,d\}} \Tr_{\C}(w e)$.
Since for $e \in \{a,b,c\}$ we also proved that $\Tr_{\D,\alpha}(w e) = \Tr_{\C}(w e)$ it follows that $\Tr_{\D,\alpha}(w d) = \Tr_{\C}(w d)$.
Hence, as above, $\Tr_{\D,\alpha}(w d^k) = \Tr_{\C}(w d^k)$ for all $k \ge 1$.
Finally, if $w \notin {(a+b)}^* \cdot (c^* + d^*)$ then $\Tr_{\D,\alpha}(w) = 0 = \Tr_{\C}(w)$.

For the converse, assume that $\A$ is not universal.
Then there is $w \in {\{a,b\}}^*$ with $\Prob_\A(w) < \frac12$.
Let $\alpha$ be a trace-based strategy.
Then we have:
\begin{align*}
\Tr_{\D, \alpha}(w c)
 & = \frac{1}{4^{|w|}} \sum_{q \in F} \dis_\A(w)(q) \cdot \alpha(w, q)(\m_c) \cdot \frac12
  && \text{by~\eqref{eq-undec-wd}} \\
 & \le \frac{1}{4^{|w|}} \cdot \frac12 \cdot \sum_{q \in F} \dis_\A(w)(q)
  && \alpha(w, q)(\m_c) \le 1 \\
 & = \frac{1}{4^{|w|}} \cdot \frac12 \cdot \Prob_\A(w)
  && \Prob_\A(w) = \sum_{q \in F} \dis_\A(w)(q) \\
 & < \frac{1}{4^{|w|}} \cdot \frac12 \cdot \frac12
  && \text{definition of~$w$} \\
 & = \Tr_\C(w c)
  && \text{Figure~\ref{fig:undecidability_m1}}
\end{align*}
We conclude that there is no trace-based strategy~$\alpha$ with $\Tr_{\D,\alpha} = \Tr_\C$.
By Lemma~\ref{lem-trace-based-enough} there is \emph{no} strategy~$\alpha$ with $\Tr_{\D,\alpha} = \Tr_\C$.
Hence $\C \not\sqsubseteq \D$.
\end{proof}

A straightforward reduction from~${\sf MDP\sqsubseteq MDP}$ now establishes:

\begin{thm}%
\label{theo-mutual-refinement}
The problem that, given two MDPs~$\D$ and $\E$, asks whether $\D \sqsubseteq \E$ and $\E \sqsubseteq \D$ is undecidable.
\end{thm}

\begin{proof}
We give a reduction from the problem ${\sf MDP \sqsubseteq MDP}$. 
Given two MDPs~$\D=\tuple{Q,\mu_0,\Label,\delta}$ and $\E=\tuple{Q',\mu'_0,\Label,\delta'}$,
we construct two MDPs called $\D + \E$ and $\E_2$ such that
$\D \sqsubseteq \E$ if, and only if, $\E_2 \sqsubseteq \D + \E$ and $\D + \E \sqsubseteq \E_2$.
See Figure~\ref{fig-mutual-refinement} for an illustration of the construction.
\begin{figure}
\centering

\tikzstyle{BoxStyle} = [draw, circle, fill=black, scale=0.4,minimum width = 1pt, minimum height = 1pt]
\begin{tikzpicture}[xscale=.6,>=latex',shorten >=1pt,node distance=3cm,on grid,auto]

 \node[label,left]  at (-2,0) {the MDP~$\D+\E$};
 \node[label,right]  at (12,0) {the MDP~$\E_2$};
 \node[state,initial,initial text={}] (p0) at (0,0) {$p_0$};
 \node[state,initial,initial where=right,initial text={}] (q0) at (10,0) {$q_0$};
 \node[state,rectangle,minimum height=40,minimum width=30] (D) at (5,1) {$\D$};
 \node[state,rectangle,minimum height=40,minimum width=30] (E) at (5,-1) {$\E$};
 \path[->] (p0) edge node [pos=0.6, above] {$\#$} (D);
 \path[->] (p0) edge node [pos=0.6, below] {$\#$} (E);
 \path[->] (q0) edge node [pos=0.6, below] {$\#$} (E);
 \node[BoxStyle,label={left,xshift=1mm,yshift=2mm}:{{$\m_\D$}}] at (2.0,0.4){};
 \node[BoxStyle,label={left,xshift=1mm,yshift=-2mm}:{{$\m_\E$}}] at (2.0,-0.4){};
 \node[BoxStyle,label={right,xshift=-1mm,yshift=-2mm}:{{$\m$}}] at (8,-0.4){};



%
\end{tikzpicture}

\caption{The construction of MDPs $\D + \E$ and $\E_2$ in the proof of Theorem~\ref{theo-mutual-refinement}.}%
\label{fig-mutual-refinement}
\end{figure}

We first construct the MDP $\D + \E$ by simply having a copy of each MDP,
adding a new label~$\#$ and a new state~$p_0$.
The  initial distribution of~$\D + \E$ is the Dirac distribution on~$p_0$, where there are
two available moves~$\m_{\D}$ and~$\m_{\E}$.
Let
$\m_{\D}(\#,q)=\mu_0(q)$ for all~$q\in Q$ and  $\m_{\D}(\#,q)=0$ otherwise;
and let  $\m_{\E}(\#,q)=\mu'_0(q)$ for all~$q\in Q'$ and  $\m_{\D}(\#,q)=0$ otherwise.
We see that $\D + \E$ always starts by generating label~$\#$ with probability~$1$;
next, using memory, $\D + \E$  can commit to either simulating~$\D$ or~$\E$.

We construct~$\E_2$ from~$\E$ as follows.
We extend the set of labels with~$\#$, and the set~$Q'$ of states with a new state~$q_0$.
The initial distribution of~$\E_2$ is the Dirac distribution on~$q_0$, where there is only one
available move~$\m$ such that $\m(\#,q)=\mu'_0(q)$.
We see that $\E_2$ always starts by generating label~$\#$ with probability~$1$, and
then simply behaves as~$\E$.

We now argue that $\D \sqsubseteq \E$ if, and only if, $\E_2 \sqsubseteq \D + \E$ and $\D + \E \sqsubseteq \E_2$.
This follows from three simple observations:
\begin{itemize}
	\item The relation $\E_2 \sqsubseteq \D + \E$ always holds. Strategies of~$\D + \E$ can choose to simulate~$\E_2$
	by playing~$\m_{\E}$ with probability~$1$.
	\item If $\D \sqsubseteq \E$ then $\D + \E \sqsubseteq \E_2$:
	a strategy~$\gamma$ of~$\D + \E$, in the first step, plays $\m_{\D}$ and~$\m_{\E}$ with probabilities $\gamma(p_0)(\m_{\D})$ and
	$\gamma(p_0)(\m_{\E})$. Next, it follows a strategy~$\alpha$ for the copy of~$\D$ and a strategy~$\beta$ for the copy of~$\E$.
	Since~$\D \sqsubseteq \E$, there exists some strategy~$\beta'$ of~$\E$ such that $\Tr_{\D,\alpha}=\Tr_{\E,\beta'}$.
    Let $\gamma_2$ be the \emph{generalized} (recall the definition preceding Lemma~\ref{lem-non-general-enough}) strategy for~$\E_2$ that first plays~$\m$, and then plays~$\beta'$ with probability~$\gamma(p_0)(\m_{\D})$
	and~$\beta$ with probability $\gamma(p_0)(\m_{\E})$.
	Then $\Tr_{\D+\E,\gamma} = \Tr_{\E_2,\gamma_2}$.
    By Lemma~\ref{lem-non-general-enough} there is a (non-generalized) strategy, $\gamma_2'$ such that $\Tr_{\E_2,\gamma_2'} = \Tr_{\E_2,\gamma_2} = \Tr_{\D+\E,\gamma}$.
    Thus  $\D + \E \sqsubseteq \E_2$.

	\item If $\D + \E \sqsubseteq \E_2$ then $\D \sqsubseteq \E$:
	consider a strategy~$\alpha$ of~$\D$. Construct strategy~$\alpha'$ of~$\D+\E$ such that
	$\alpha'(p_0)(\m_{\D})=1$ and $\alpha'(p_0\#\rho)=\alpha(\rho)$ for all paths~$\rho$.
	Since~$\D + \E \sqsubseteq \E_2$, there must be some strategy~$\beta'$ such that $\Tr_{\D+\E,\alpha'}=\Tr_{\E_2,\beta'}$.
	For the strategy~$\beta$ where $\beta(\rho)=\beta'(q_0\#\rho)$ for all paths~$\rho$, we have $\Tr_{\D,\alpha}=\Tr_{\E,\beta}$, and thus
	$\D \sqsubseteq \E$. \qedhere
\end{itemize}
\end{proof}

\section{Decidability for Memoryless Strategies}%
\label{sec-memoryless-strategies}
Given two MCs~$\C_1$ and~$\C_2$, the (symmetric) trace-equivalence relation~$\C_1 \sqsubseteq \C_2$
is polynomial-time decidable~\cite{Tzeng96}.
An MDP~$\D$ under a  memoryless strategy~$\alpha$ induces a finite MC~$\D(\alpha)$, and thus
once a  memoryless strategy is fixed for the MDP,
its relation to another given MC in the trace-equivalence relation~$\sqsubseteq$ can be decided in~{\sf P}.
Theorems~\ref{theo:MCMDPPM}~and~\ref{theo:MDPPM} provide tight complexity bounds of the trace-refinement problems for
MDPs that are restricted to use only pure memoryless strategies.
In Theorems~\ref{thm-reduction-to-ExThR}~and~\ref{thm-hardness-for-NMF} we establish bounds on the complexity of the problem when randomization is allowed for memoryless strategies.

\subsection{Pure  Memoryless Strategies}%
\label{subsec:purememoryless}
In this subsection, we show that the problems ${\sf MC\sqpm MDP_{}}$  and~${\sf MDP_{}\sqpm MDP_{}}$
are \mbox{{\sf NP}-complete} and $\Pi^p_2$-complete, respectively.
%
%
The hardness results are by reductions from the \emph{subset-sum problem}
and a variant of the \emph{quantified subset-sum} problem.

Given a set~$\{s_1, s_2,\dots,s_n\}$ of natural numbers and~$N\in \nat$,
the subset-sum problem asks whether there exists a subset~$S\subseteq \{s_1, \dots,s_n\}$ such that
$\sum_{s\in S}s=N$. The subset-sum problem is known to be {\sf NP}-complete~\cite{Cormen01}.
The  quantified version of subset sum   is a game between a \emph{universal player}
and an \emph{existential player}.
Given~$k,N\in \nat$ and two sets
$\{s_1, s_2,\dots,s_n\}$  and $\{t_1, t_2,\dots,t_m\}$ of natural numbers, the game is played turn-based for~$k$ rounds.
In each round~$i$ ($1\leq i \leq k$), the universal player first chooses~$S_i\subseteq \{s_1, \dots,s_n\}$
and then the existential player chooses~$T_i\subseteq \{t_1, \dots,t_m\}$.
The existential player wins if and only if
\[\sum_{s\in S_1}s+\sum_{t\in T_1}t+ \cdots +\sum_{s\in S_k}s+\sum_{t\in T_k}t =N.\]
The \emph{quantified subset-sum} problem is to check whether the existential player has a winning strategy.
The problem is known to be {\sf PSPACE}-complete~\cite{FearnleyJ15}.
The proof therein implies  that  the variant of the problem with a fixed number~$k$ of rounds is~$\Pi^p_{2k}$-complete.

\begin{thm}%
\label{theo:MCMDPPM}
The problem~${\sf MC\sqpm MDP_{}}$ is \mbox{{\sf NP}-complete}.
\end{thm}

\begin{proof}
Membership of ${\sf MC\sqpm MDP_{}}$ in {\sf NP} is obtained as follows.
Given an MC~$\C$ and an MDP~$\D$, the polynomial-time verifiable witness of~$\C\sqsubseteq \D$
is a pure memoryless strategy~$\alpha$ for~$\D$.
Once~$\alpha$ is fixed, then~$\C \sqsubseteq \D(\alpha)$ can be decided in~{\sf P}.

To establish {\sf NP}-hardness of ${\sf MC\sqpm MDP_{}}$,
consider an instance of subset sum, i.e., a  set~$\{s_1, \dots,s_n\}$  and~$N\in \nat$.
We can assume without loss of generality that $N \le P$, where $P = s_1+\cdots+s_n$.
We construct
an MC~$\C$ and an MDP~$\D$ such that
there exists~$S\subseteq \{s_1, \dots,s_n\}$ with
$\sum_{s\in S}s=N$ if and only if~$\C \sqsubseteq \D$ when~$\D$ uses only pure memoryless strategies.

\begin{figure}[t]
\begin{minipage}[b]{0.45\linewidth}
\centering
\centering
\begin{tikzpicture}[xscale=.6,>=latex',shorten >=1pt,node distance=3cm,on grid,auto]

\node[label]  at (2,2.7) {the MC~$\C$};
\node[state,initial,initial where=left,initial text={}] (q0) at (2,2) {$q_{0}$};
\node[state] (qb) at (0,0) {$q_b$};
\node[state] (qc) at (4,0) {$q_c$};

 \path[->] (q0) edge  node [midway, left] {$\frac{N}{P},a$} (qb);
 \path[->] (q0) edge  node [midway, right] {$1-\frac{N}{P},a$} (qc);

 \path[->] (qc) edge  [out=15,in=-15,looseness=8] node [midway, right] {$1, c$} (qc);
 \path[->] (qb) edge [out=165,in=195,looseness=8]   node [midway, left]{$1, b$} (qb);

\end{tikzpicture}

\end{minipage}
\hspace{0.5cm}
\begin{minipage}[b]{0.45\linewidth}
\centering
\centering

\tikzstyle{BoxStyle} = [draw, circle, fill=black, scale=0.4,minimum width = 1pt, minimum height = 1pt]
\begin{tikzpicture}[xscale=.6,>=latex',shorten >=1pt,node distance=3cm,on grid,auto]

\node[label]  at (2.8,2.7) {the MDP~$\D$};

 \node[state] (u1) at (0,2) {$s_{1}$};
 \node[label] at (2.9,2) {$\cdots$};
 \node[state] (un) at (6,2) {$s_{n}$};

 \node[BoxStyle,label=left:{{$\m_{1,b}$}}] at (.35,1.3){};
 \node[BoxStyle,label=right:$\m_{1,c}$] at (.95,1.6){};
 \node[BoxStyle,label=left:{{$\m_{n,b}$}}] at (5.05,1.6){};
 \node[BoxStyle,label=right:$\m_{n,c}$] at (5.65,1.3){};

 \node[state] (qb) at (1,0) {$s_b$};
\node[state] (qc) at (5,0) {$s_c$};


 \path[->] (qc) edge  [out=15,in=-15,looseness=8] node [midway, right] {$1, c$} (qc);
 \path[->] (qb) edge [out=165,in=195,looseness=8]   node [midway, left]{$1, b$} (qb);

\path[->] (u1) edge node  [very near end,left] {$a$} (qb);
\path[->] (u1) edge node  [very near end,above] {$a$} (qc);
\path[->] (un) edge node [very near end,above] {$a$} (qb);
\path[->] (un) edge node [very near end,right] {$a$} (qc);

\end{tikzpicture}

\end{minipage}
\caption{The MC~$\C$  and the MDP~$\D$ in the reduction for {\sf NP}-hardness
of~${\sf MC\sqm MDP_{}}$.}\label{fig:reductionfromSubsset}
\end{figure}

The MC~$\C$ is shown in Figure~\ref{fig:reductionfromSubsset} on the left.
The initial distribution is the Dirac distribution on~$q_0$;
$\C$ generates traces in~$ab^+$ with probability~$\frac{N}{P}$ and
traces in~$ac^+$ with probability~$1-\frac{N}{P}$.

The MDP~$\D$ is shown in Figure~\ref{fig:reductionfromSubsset} on the right.
For all states~$s_i$, two moves~$\m_{i,b}$ and~$\m_{i,c}$ are available,
the Dirac distributions on~$(a,s_b)$ and~$(a,s_c)$.
The states~$s_{b},s_c$ emit only the single labels~$b$ and~$c$.
The initial  distribution~$\mu_0$ is such that
$\mu_0(s_i)=\frac{s_i}{P}$ for all~$1\leq i \leq n$.
Intuitively, choosing~$b$ in~$s_i$ simulates the membership of~$s_i$ in~$S$ by adding~$\frac{s_i}{P}$
to the probability of generating~$ab^{+}$.

For a pure strategy~$\alpha$ for~$\D$,
let~$S_{\alpha}$ be the set of states~$s_i$ where~$\alpha(s_i)=m_{i,b}$. Then,
 $\Tr_{D}(ab^{+})=\sum_{s\in S_{\alpha}} \frac{s}{P}$ and
$\Tr_{D}(ac^{+})=1-Tr_{D}(ab^{+})$.
Hence
$\C \sqsubseteq \D$ holds if and only if there exists a strategy~$\alpha$ for~$\D$
such that
$\sum_{s\in S_{\alpha}} \frac{s}{P}=\frac{N}{P}$.
It implies that
the instance of subset problem is positive, meaning that
there exists a subset~$S\subseteq \{s_1, s_2,\dots,s_n\}$ such that
$\sum_{s\in S}s=N$, if and only if $\C \sqsubseteq \D$ when $\D$  uses  only pure memoryless strategies.
The {\sf NP}-hardness results follows.
\end{proof}

In the next theorem, we show that  ${\sf MDP_{}\sqpm MDP_{}}$ is  $\Pi^p_2$-complete.
The hardness is by reduction from the quantified subset-sum problem with  $k=1$ (one alternation).
\begin{thm}%
\label{theo:MDPPM}
The problem~${\sf MDP_{}\sqpm MDP_{}}$
is $\Pi^p_2$-complete.
\end{thm}

\begin{proof}

Membership of  ${\sf MDP_{}\sqpm MDP_{}}$ in $\Pi^p_2$ is obtained as follows.
Let~$\D$ and $\E$ be two MDPs. To check~$\E \sqsubseteq \D$,
for all  pure memoryless strategies~$\beta$ of~$\E$
one can guess a polynomial-time verifiable witness~$\alpha$, a strategy of $\D$.
Once $\alpha$ and $\beta$ are fixed in $\D$ and $\E$ respectively,
checking~$\E(\beta) \sqsubseteq \D(\alpha)$
can be done in~{\sf P}.

 To establish the hardness,
consider an instance of quantified subset sum, i.e., $N\in \nat$
and two sets~$\{s_1,\dots,s_n\}$ and $\{t_1,\dots,t_m\}$.
We construct MDPs~$\E_{univ}$ and $\E_{exist}$ such that
the existential player wins in one round  if and only if $\E_{univ} \sqsubseteq \E_{exist}$ holds,
where the MDPs  use only pure memoryless strategies.

Let $P=s_1+\cdots+s_n$ and $R=t_1+\cdots+t_m$.
Pick a small real number $0<x<1$ so that $0 < x P, x R, x N < 1$.
Pick real numbers $0\leq y_1,y_2 < 1$ such that
$y_1+x N<1$ and $y_1+x N=y_2+x R$.


The MDPs~$\E_{univ}$ and $\E_{exist}$ have  symmetric constructions.
The MDP~$\E_{univ}$ simulates  choices of the universal player
and is drawn in Figure~\ref{fig:reductionfromQGSS} on the left.
For all states~$s_i$, two moves~$\m_{i,b}$ and~$\m_{i,c}$ are available,
the Dirac distributions on~$(a,s_b)$ and~$(a,s_c)$.
The initial distribution~$\mu_0$ for~$\E_{univ}$ is such that
$\mu_0(s_y)=\frac{1}{2}y_1$ and $\mu_0(s_r)=1-\frac{1}{2}(xP+y_1)$, and
 $\mu_0(s_i)=\frac{1}{2}xs_i$ for all~$1\leq i \leq n$.
The MDP $\E_{exist}$ simulates  choices of the existential player
and is drawn in Figure~\ref{fig:reductionfromQGSS} on the right.
For all states~$t_i$, two moves~$\m_{i,b}$ and~$\m_{i,c}$ are available,
the Dirac distributions on~$(a,t_b)$ and~$(a,t_c)$, similar to~$\E_{univ}$.
The initial distribution~$\mu'_0$ for $\E_{exist}$ is such that
$\mu'_0(t_y)=\frac{1}{2}y_2$ and $\mu'_0(t_r)=1-\frac{1}{2}(xR+y_2)$,
and $\mu'_0(t_j)=\frac{1}{2}xt_j$
for all~$1\leq j \leq m$.
Choosing~$b$ in a set of states~$s_i$ by the
universal player is responded by choosing~$c$ in a right set of states~$t_j$
by the existential player such that the
probabilities of emitting~$ab^{+}$ in the MDPs are equal.

\begin{figure}[t]
\begin{minipage}[b]{0.45\linewidth}
\centering
\centering

\tikzstyle{BoxStyle} = [draw, circle, fill=black, scale=0.4,minimum width = 1pt, minimum height = 1pt]
\begin{tikzpicture}[xscale=.6,>=latex',shorten >=1pt,node distance=3cm,on grid,auto]

\node[label]  at (2.8,2.7) {the MDP~$\E_{univ}$};

\node[state] (sr) at (7.5,0) {$s_{r}$};
 \node[state] (u1) at (0,2) {$s_{1}$};
 \node[label] at (2.9,2) {$\cdots$};
 \node[state] (un) at (6,2) {$s_{n}$};
 \node[state] (sy) at (-1.5,0) {$s_{y}$};

 \node[BoxStyle,label=left:{{$\m_{1,b}$}}] at (.35,1.3){};
 \node[BoxStyle,label=right:$\m_{1,c}$] at (.95,1.6){};
 \node[BoxStyle,label=left:{{$\m_{n,b}$}}] at (5.05,1.6){};
 \node[BoxStyle,label=right:$\m_{n,c}$] at (5.65,1.3){};

 \node[state] (qb) at (1,0) {$s_b$};
\node[state] (qc) at (5,0) {$s_c$};


 \path[->] (qc) edge  [out=-75,in=-105,looseness=4] node [midway, right] {$1, c$} (qc);
 \path[->] (qb) edge [out=-75,in=-105,looseness=4]   node [midway, right]{$1, b$} (qb);
\path[->] (sy) edge node  [midway,below] {$a$} (qb);
\path[->] (sr) edge node  [midway,below] {$a$} (qc);
\path[->] (u1) edge node  [very near end,left] {$a$} (qb);
\path[->] (u1) edge node  [very near end,above] {$a$} (qc);
\path[->] (un) edge node [very near end,above] {$a$} (qb);
\path[->] (un) edge node [very near end,right] {$a$} (qc);

\end{tikzpicture}

\end{minipage}
\hspace{0.5cm}
\begin{minipage}[b]{0.45\linewidth}
\centering
\centering

\tikzstyle{BoxStyle} = [draw, circle, fill=black, scale=0.4,minimum width = 1pt, minimum height = 1pt]
\begin{tikzpicture}[xscale=.6,>=latex',shorten >=1pt,node distance=3cm,on grid,auto]

\node[label]  at (2.8,2.7) {the MDP~$\E_{exist}$};

\node[state] (sr) at (7.5,0) {$t_{r}$};
 \node[state] (u1) at (0,2) {$t_{1}$};
 \node[label] at (2.9,2) {$\cdots$};
 \node[state] (un) at (6,2) {$t_{m}$};
 \node[state] (sy) at (-1.5,0) {$t_{y}$};

 \node[BoxStyle,label=left:{{$\m_{1,b}$}}] at (.35,1.3){};
 \node[BoxStyle,label=right:$\m_{1,c}$] at (.95,1.6){};
 \node[BoxStyle,label=left:{{$\m_{m,b}$}}] at (5.05,1.6){};
 \node[BoxStyle,label=right:$\m_{m,c}$] at (5.65,1.3){};

 \node[state] (qb) at (1,0) {$t_b$};
\node[state] (qc) at (5,0) {$t_c$};


 \path[->] (qc) edge  [out=-75,in=-105,looseness=4] node [midway, right] {$1, c$} (qc);
 \path[->] (qb) edge [out=-75,in=-105,looseness=4]   node [midway, right]{$1, b$} (qb);
\path[->] (sy) edge node  [midway,below] {$a$} (qb);
\path[->] (sr) edge node  [midway,below] {$a$} (qc);
\path[->] (u1) edge node  [very near end,left] {$a$} (qb);
\path[->] (u1) edge node  [very near end,above] {$a$} (qc);
\path[->] (un) edge node [very near end,above] {$a$} (qb);
\path[->] (un) edge node [very near end,right] {$a$} (qc);

\end{tikzpicture}

\end{minipage}
\caption{The MDPs~$\E_{univ}$ and $\E_{exist}$ in the reduction for $\Pi^p_2$-hardness
of~${\sf MDP_{}\sqpm MDP_{}}$.}\label{fig:reductionfromQGSS}
\end{figure}


For a pure strategy~$\alpha$ of~$\E_{univ}$,
let~$S_{\alpha}$ be the set of states~$s_i$ where~$\alpha(s_i)=m_{i,b}$.
We therefore have
 $\Tr_{\E_{univ}}(ab^{+})=\frac{1}{2} y_1+\frac{1}{2}\sum_{s\in S_{\alpha}} xs$.
For a pure   strategy~$\beta$ of~$\E_{exist}$,
let~$T_{\beta}$ be the set of states~$t_j$ where~$\beta(t_j)=m_{i,c}$. Then,
$\Tr_{\E_{exist}}(ac^{+})=1-\frac{1}{2}(xR+y_2)+\frac{1}{2}\sum_{t\in T_{\beta}} xt$.
It implies that
$\Tr_{\E_{exist}}(ab^{+})=\frac{1}{2}(xR+y_2)-\frac{1}{2}\sum_{t\in T_{\beta}} xt$.

Since $y_1+x N=y_2+x R$, to achieve  $\Tr_{\E_{univ}}= \Tr_{\E_{exist}}$  the equality~$\sum_{s\in S_{\alpha}}s=N-\sum_{t\in T_{\beta}}t$ must be guaranteed.
It shows that
the existential player wins in one round, meaning that
for all subsets~$S\subseteq \{s_1, s_2,\dots,s_n\}$
there exists a subset~$T\subseteq \{t_1, t_2,\dots,t_m\}$
such that
$\sum_{s\in S}s+\sum_{t\in T}t=N$, if and only if for all pure and memoryless
strategies~$\alpha$ of~$\D$
there exists some pure and memoryless strategy~$\beta$ for~$\E$ such that
  $\Tr_{\E_{univ}}= \Tr_{\E_{exist}}$.
The $\Pi^p_2$-hardness result follows.
\end{proof}

\subsection{Memoryless Strategies}\label{subsec:memoryless}

In this subsection, we provide upper and lower complexity bounds for the problem~${\sf MC\sqm MDP_{}}$:
a reduction to the existential theory of the reals
and a reduction from nonnegative matrix factorization.

A formula of the \emph{existential theory of the reals} is of the form
 $\exists x_1 \dots \exists x_m~R(x_1, \dots, x_n)$, where $R(x_1, \dots, x_n)$ is a boolean combination of comparisons of the form
 $p(x_1, \dots, x_n) \sim 0$, where $p(x_1, \dots, x_n)$ is a multivariate polynomial and
  $\mathord{\sim} \in \{ \mathord{<}, \mathord{>}, \mathord{\le}, \mathord{\ge}, \mathord{=}, \mathord{\ne} \}$.
The validity of closed formulas (i.e., when $m=n$) is decidable in {\sf PSPACE}~\cite{Can88,Renegar92},
 and is not known to be {\sf PSPACE}-hard.

\begin{thm}%
\label{thm-reduction-to-ExThR}
The problem~${\sf MC\sqm MDP_{}}$ is polynomial-time reducible to the existential theory of the reals,
hence in {\sf PSPACE}.
\end{thm}

Given an MC~$\C = \tuple{Q,\mu_{0},\Label,\delta}$,
to each label~$a \in \Label$ we associate  a \emph{transition matrix} $\Delta(a) \in {[0,1]}^{Q \times Q}$ with
$\Delta(a)[q,q'] = \delta(q)(a,q')$.
We view subdistributions~$\mu_{0}$ over states as row vectors $\mu_{0} \in {[0,1]}^Q$.
We denote column vectors in boldface; in particular,
$\vec{1} \in {\{1\}}^{Q}$ and~$\vec{0} \in {\{0\}}^{Q}$ are column vectors all whose
entries are $1$ and~$0$, respectively.
We build on~\cite[Proposition~10]{14KW-ICALP} which reads---translated to our framework---as follows:

\begin{prop}%
\label{prop-Bjoern}
Let $\C_1 = \tuple{Q_1,\mu_{0},\Label,\delta}$ and $\C_2 = \tuple{Q_2,\mu'_{0},\Label,\delta'}$
be MCs with $Q$ as the disjoint union of $Q_1, Q_2$.
Then $\Tr_{\C_1} = \Tr_{\C_2}$ if and only if there exists a  matrix $F \in \Reals^{Q \times Q}$
such that
\begin{itemize}
\item the first row of~$F$ equals $(\mu_0, -\mu'_0)$,
\item $F \vec{1} = \vec{0}$,
\end{itemize}
and, moreover, for all labels~$a \in \Label$ there exist matrices $M(a) \in \Reals^{Q \times Q}$ such that
\[
F \begin{pmatrix} \Delta(a) \ & \ 0 \\ 0 \ & \ \Delta'(a) \end{pmatrix}
= M(a) F
\]
where  $\Delta(a), \Delta(a)'$ are the transition matrices of~$\C_1$ and~$\C_2$ for the label~$a$.
\end{prop}

With this at hand we prove Theorem~\ref{thm-reduction-to-ExThR}:

\begin{proof}[Proof of Theorem~\ref{thm-reduction-to-ExThR}]
Let $\C = \tuple{Q_1,\mu_{0},\Label,\delta}$ be an MC and $\D = \tuple{Q_2,\mu'_{0},\Label,\delta'}$ be an MDP
 with $Q$ as the disjoint union of $Q_1, Q_2$.
A memoryless strategy~$\alpha$ of~$\D$ can be characterized by numbers $x_{q,\m} \in [0,1]$
where $q \in Q_2$ and $\m \in \moves(q)$, such that $x_{q,\m} = \alpha(q)(\m)$.
We have
$\sum_{\m \in \moves(q)} x_{q,\m} = 1$ for all states~$q$.
\newcommand{\barx}{\overline{x}}%
We write $\barx$ for the collection ${(x_{q,\m})}_{q \in Q_2, \ \m \in \moves(q)}$, and $\alpha(\barx)$ for the memoryless strategy characterized by~$\barx$.
We have:
\begin{align*}
\C \sqm \D 
& \ \Longleftrightarrow \ \exists\, \text{memoryless strategy }\alpha: \Tr_{\C} = \Tr_{\D,\alpha}
 && \text{definition} \\
& \ \Longleftrightarrow \ \Cond
 && \text{Proposition~\ref{prop-Bjoern}},
\end{align*}
where $\Cond$ is the following condition:
\begin{quote}
There exist
\begin{itemize}
\item $x_{q,\m} \in [0,1]$ for all $q \in Q_2$ and all $\m \in \moves(q)$
\item matrices $M(a) \in \Reals^{Q \times Q}$ for all labels~$a \in \Label$,
\item a matrix $F \in \Reals^{Q \times Q}$
\end{itemize}
such that
\begin{itemize}
\item  $\sum_{\m \in \moves(q)} x_{q,\m} = 1$ for all $q \in Q_2$,
\item
the first row of~$F$ equals $(\mu_0, -\mu'_0)$,
\item $F \vec{1} = \vec{0}$,
\item for all labels~$a \in \Label$,
\[
F \begin{pmatrix} \Delta(a) \ & \ 0 \\ 0 \ & \ \Delta'(a) \end{pmatrix}
= M(a) F
\]
where $\Delta(a),\Delta'(a)$ are the transition matrices of~$C$ and
the finite MC~$\D(\alpha(\barx))$ induced by~$\D$ under the strategy~$\alpha(\barx)$.
\end{itemize}
\end{quote}
This condition~$\Cond$ is a closed formula in the existential theory of the reals.
\end{proof}

Given a nonnegative matrix $M \in \mathbb{R}^{n \times m}$, a \emph{nonnegative factorization} of $M$ with \emph{inner dimension~$r$}
is a decomposition of the form $M = A \cdot W$ where $A \in \mathbb{R}^{n \times r}$ and $W \in \mathbb{R}^{r \times m}$
are nonnegative matrices (see~\cite{CohenR93,Vavasis09,AroraGKM12} for more details).
The \emph{NMF problem} asks, given a nonnegative matrix $M \in \mathbb{R}^{n \times m}$ and a number $r \in \nat$,
whether there exists a factorization $M = A \cdot W$ with nonnegative matrices $A \in \mathbb{R}^{n \times r}$ and $W \in \mathbb{R}^{r \times m}$.
The NMF problem is known to be {\sf NP}-hard~\cite{Vavasis09}.


\begin{thm}%
\label{thm-hardness-for-NMF}
The NMF problem is polynomial-time reducible to~${\sf MC\sqm MDP_{}}$,
hence ${\sf MC\sqm MDP_{}}$ is {\sf NP}-hard.
\end{thm}

\begin{proof}

%
To establish the reduction,
consider an instance of the NMF problem, a nonnegative matrix $M \in \mathbb{R}^{n \times m}$ and a number~$r \in \nat$.
We construct an MC~$\C$ and an MDP~$\D$ such that the NMF instance is a yes-instance
if and only if $\C \sqsubseteq \D$
where~$\D$ is restricted to use only memoryless strategies.

We can assume, without loss of generality~\cite[Section~3]{CohenR93}, that~$M$ is a stochastic matrix, that is
$\sum\limits_{j=1}^m M[i,j]=1$ for all rows~$1\leq i\leq n$.
Also by~\cite[Section~3]{CohenR93} we know that
there exists a nonnegative factorization of $M$  with inner dimension~$r$ if and only if
there exist two stochastic matrices $A \in \mathbb{R}^{n \times r}$ and $W \in \mathbb{R}^{r \times m}$ such that
$M=A \cdot W$.

The transition probabilities in the MC~$\C$ encode the entries of matrix~$M$.
The initial distribution of the MC is the Dirac distribution on~$q_{\inn}$; see \figurename~\ref{fig:reductionfromNMFMC}.
\begin{figure}
\centering
\centering
\begin{tikzpicture}[xscale=.8,>=latex',shorten >=1pt,node distance=3cm,on grid,auto]

 \node[state,initial,initial text={}] (qin) at (0,2) {$q_{\inn}$};

 \node[state] (q1) at (2,4) {$q_{1}$};
 \node[label] at (1,2.1) {$\vdots$};
 \node[state] (qn) at (2,0) {$q_{n}$};

\node[state] (qfi) at (8,2) {$q_{\fin}$};


 \path[->] (qin) edge  node [midway, left] {$\frac{1}{n},a_1$} (q1);
 \path[->] (qin) edge  node [midway, left] {$\frac{1}{n},a_n$} (qn);

 \path[->] (q1) edge [bend left=15] node [midway, right,xshift=2mm] {$M[1,1],b_1$} (qfi);
 \node[label] at (4.8,3.3) {$\vdots$};
 \path[->] (q1) edge [bend right=15] node [midway, left, xshift=-2mm,yshift=-1mm,] {$M[1,m],b_m$} (qfi);

 \path[->] (qn) edge [bend left=15] node [midway, left, xshift=-2mm,yshift=1mm] {$M[n,1],b_1$} (qfi);
 \node[label] at (4.8,.8) {$\vdots$};
 \path[->] (qn) edge [bend right=15] node [midway, right,xshift=2mm,yshift=-1mm] {$M[n,m],b_m$} (qfi);

\path[->] (qfi) edge [loop right] node [right, midway] {$c,1$}  (qfi);

\end{tikzpicture}

\caption{The MC~$\C$ of the reduction from NMF to~${\sf MC\sqm MDP_{}}$.}\label{fig:reductionfromNMFMC}
\end{figure}
There are $n+m+1$ labels~$a_1,\dots,a_n,b_1,\dots,b_m,c$.
The transition in~$q_{\inn}$ is the uniform distribution over~$\{(a_i,q_i) \mid 1\leq i \leq n\}$.
In each state~$q_i$, each  label~$b_j$ is emitted with probability~$M[i,j]$, and a transition to~$q_{\fin}$
is taken.
In state~$q_{\fin}$ only~$c$ is emitted.
Observe that for all $1\leq i\leq n$ and  $1\leq j\leq m$ we have
$\Tr_{\C}(a_i)=\frac{1}{n}$ and $\Tr_{\C}(a_i\cdot b_j \cdot c^{*})=\frac{1}{n}M[i,j]$.

The initial distribution of the MDP~$\D$ is the uniform distribution over~$\{p_1,\dots,p_n\}$;
see \figurename~\ref{fig:reductionfromNMFMDP}.
In each~$p_i$ (where~$1\leq i\leq n$), there are $r$ moves~$\m_{i,1}, \m_{i,2},\dots,\m_{i,r}$
where~$\m_{i,k}(a_i,\ell_k)=1$ and $1\leq k \leq r$.
In each~$\ell_k$, there are $m$ moves~$\m'_{k,1}, \m'_{k,2},\dots,\m'_{k,m}$ where
$\m'_{k,j}(b_j,p_{\fin})=1$ where  $1\leq j\leq m$.
In state~$p_{\fin}$, only~$c$ is emitted.
The probabilities of choosing the move~$m_{i,k}$ in~$p_i$ and
choosing~$m'_{k,j}$ in~$\ell_k$ simulate the entries of~$A[i,k]$ and~$W[k,j]$.

\begin{figure}
\centering
\centering

\tikzstyle{BoxStyle} = [draw, circle, fill=black, scale=0.4,minimum width = 1pt, minimum height = 1pt]
\begin{tikzpicture}[xscale=.6,>=latex',shorten >=1pt,node distance=3cm,on grid,auto]

 \node[state,initial,initial text={$\frac{1}{n}$}] (p1) at (0,4) {$p_{1}$};
 \node[label] at (0,2.1) {$\vdots$};
 \node[state,initial,initial text={$\frac{1}{n}$}] (pn) at (0,0) {$p_{n}$};

 \node[BoxStyle,label={{$\m_{1,1}$}}] at (1.3,4){};
 \node[BoxStyle,label={left,yshift=-1mm}:$\m_{1,r}$] at (.7,3.3){};

 \node[BoxStyle,label={below,xshift=1mm}:$\m_{n,r}$] at (1.3,0){};
 \node[BoxStyle,label=left:$\m_{n,1}$] at (.7,.7){};

 \node[state] (l1) at (4,4) {$\ell_{1}$};
 \node[label] at (4,2.1) {$\vdots$};
 \node[state] (lk) at (4,0) {$\ell_{r}$};

 \node[BoxStyle, label=above:$\m'_{1,1}$] at (5.3,3.9){};
 \node[BoxStyle,label=below:$\m'_{1,r}$] at (4.9,3.4){};

 \node[BoxStyle,label={below,xshift=1mm}:$\m'_{n,r}$] at (5.3,.1){};
 \node[BoxStyle,label={above,xshift=-1mm}:$\m'_{n,1}$] at (4.9,.6){};

\node[state] (pfi) at (10,2) {$p_{\fin}$};

\path[->] (p1) edge node  [very near end] {$a_1$} (l1);
\path[->] (p1) edge node  [very near end,right] {$a_1$} (lk);

\path[->] (pn) edge node [very near end,right] {$a_n$} (l1);
\path[->] (pn) edge node [very near end,above] {$a_n$} (lk);

 \path[->] (l1) edge [bend left=15]  node [near end, above] {$b_1$} (pfi);
 \node[label] at (6.8,3.3) {$\vdots$};
 \path[->] (l1) edge [bend right=15] node [near end, above] {$b_m$} (pfi);
 \path[->] (lk) edge [bend left=15] node [near end, below] {$b_1$} (pfi);
 \node[label] at (6.8,3.3) {$\vdots$};
 \path[->] (lk) edge [bend right=15] node[yshift=-1mm] [near end, below] {$b_m$} (pfi);

\path[->] (pfi) edge [loop right] node [right, midway] {$c,1$}  (pfi);
\end{tikzpicture}

\caption{The MDP~$\D$ of the reduction from NMF to~${\sf MC\sqm MDP_{}}$.}\label{fig:reductionfromNMFMDP}
\end{figure}

We prove that there is a nonnegative factorization for~$M=A\cdot W$ such that
$A \in \mathbb{R}^{n \times r}$ and $W \in \mathbb{R}^{r \times m}$
if and only if $\C \sqsubseteq \D$
where~$\D$ is restricted to memoryless strategies.

Suppose $M$ has a nonnegative factorization, i.e., there are stochastic matrices $A \in \mathbb{R}^{n \times r}$ and $W \in \mathbb{R}^{r \times m}$ such that
$M=A \cdot W$.
To prove that $\C \sqsubseteq \D$, we construct a memoryless strategy~$\alpha$ such that~$\Tr_{\C}=\Tr_{\D,\alpha}$.
For all states~$q$ of~$\D$, strategy~$\alpha$ is defined by
\[
\alpha(q) =
\left\{
	\begin{array}{ll}
		d\in \dists(\moves(p_i))   & \mbox{if } q=p_i \text{ and } 1\leq i\leq n,\\
		\text{where }  d(\m_{i,k})=A[i,k] \text{ for all } 1 \leq k\leq r \\
		&\\
		d\in \dists(\moves(\ell_k))   & \mbox{if } q=\ell_k \text{ and } 1\leq k\leq r,\\
		\text{where }  d(\m'_{k,j})=W[k,j] \text{ for all } 1 \leq j\leq m \\
		&\\
		\text{the Dirac distribution on~$(c,p_{\fin})$ }  & \mbox{if } q=p_{\fin}.\\
	\end{array}
\right.
\]
The trace-probability function for~$\D$ and $\alpha$ is such that for all $1\leq i\leq n$ and all $1\leq j\leq m$,
we have  $\Tr_{\D,\alpha}(a_i)=\frac{1}{n}$,
and
\begin{align*}
\Tr_{\D,\alpha}(a_i\cdot b_j \cdot c^{*})& =\frac{1}{n} \sum\limits_{k=1}^{r}\alpha(p_i)(\m_{i,k})\cdot \alpha(\ell_k)(\m'_{k,j})\\
 & =\frac{1}{n} \sum\limits_{k=1}^{r} A[i,k]\cdot W[k,j] =\frac{1}{n}M[i,j].
\end{align*}
This gives  $\Tr_{\D,\alpha}=\Tr_{\C}$, and thus $\C \sqsubseteq \D$
where~$\D$ uses a memoryless strategy.

Conversely, suppose that there exists a memoryless strategy~$\beta$ for the MDP~$\D$ such that~$\Tr_{\C}=\Tr_{\D,\beta}$.
We exhibit a factorization $M=A \cdot W$ where $A \in \mathbb{R}^{n \times r}$ and $W \in \mathbb{R}^{r \times m}$.
For all $1\leq i \leq n$,  $1 \leq k\leq r$ and $1\leq j\leq m$, let
\[A[i,k]=\beta(p_i)(\m_{i,k}) \quad \quad \text{ and } \quad \quad W[k,j]=\beta(\ell_k)(\m'_{k,j}).\]
Since $\D$ under the strategy~$\beta$ refines~$\C$, then
for all $1\leq i\leq n$ and all $1\leq j\leq m$
\[\Tr_{\D,\beta}(a_i\cdot b_j \cdot c^{*})=\Tr_{\C}(a_i\cdot b_j \cdot c^{*})=\frac{1}{n}M[i,j].\]
Since the probability of generating~$a_i\cdot b_j \cdot c^{*}$  is
$\frac{1}{n} \sum\limits_{k=1}^{r}\beta(p_i)(\m_{i,k})\cdot \beta(\ell_k)(\m'_{k,j})$
then we have
\[\sum\limits_{k=1}^{r} A[i,k]\cdot W[k,j] =M[i,j].\]
This completes the proof.
\end{proof}

Shitov~\cite{Shitov16} recently claimed that the NMF problem
is complete for the existential theory of the reals.
This claim, combined with
Theorems~\ref{thm-reduction-to-ExThR}~and~\ref{thm-hardness-for-NMF},
implies that  ${\sf MC\sqm MDP_{}}$ is complete for
the existential theory of the reals.


\section{Bisimulation}%
\label{sec-BisimulationRefinements}
In this section we show that the problem~${\sf MDP\sqsubseteq MC}$ is in~{\sf NC}, hence in~{\sf P}.

First, in Subsection~\ref{subsec:link-trace-refinement-bisim}, we establish a link between trace refinement and a notion of bisimulation between distributions that was studied in~\cite{Jans}.

Second, in Subsection~\ref{subsec:condition_bisimilar} we give a necessary and sufficient condition for the MDPs to be bisimilar.
It resembles the properties developed in~\cite{Jans}, but we rebuild a detailed proof from scratch, as the authors were unable to verify some of the technical claims made in~\cite{Jans}.

As corollaries, we show in Subsection~\ref{subsec:algo_conp_bisimilarity} that bisimulation between two MDPs can be decided in~{\sf coNP}, improving the exponential-time result from~\cite{Jans},
and in Subsection~\ref{subsec:algo_nc_trace_refinement} that the problem~${\sf MDP\sqsubseteq MC}$ is in~{\sf NC}, hence in~{\sf P}.

\subsection{A Link between Trace Refinement and Bisimulation}%
\label{subsec:link-trace-refinement-bisim}

A \emph{local strategy} for an MDP $\D = \tuple{Q,\mu_{0},\Label,\delta}$ is a function~$\alpha : Q \to \dists(\moves)$
that maps each state~$q$ to a distribution~$\alpha(q)\in \dists(\moves(q))$ over moves in~$q$.
We call~$\alpha$ \emph{pure} if for all states~$q$ there is a move~$\m$ such that $\alpha(q)(\m) = 1$.
%
For a subdistribution~$\mu \in \subdists(Q)$, a local strategy~$\alpha$, and a label~$a \in \Label$,
define the \emph{successor} subdistribution $\Succ(\mu, \alpha, a)$ with
\[
\Succ(\mu, \alpha, a)(q')
= \sum_{q \in Q} \mu(q) \cdot \sum_{\m \in \moves(q)} \alpha(q)(\m) \cdot \m(a, q')
\]
for all $q' \in Q$.
We often view a subdistribution~$d \in \subdists(Q)$ as a row vector $d \in {[0,1]}^Q$.
For a local strategy $\alpha$ and a label~$a$,
define the \emph{transition matrix} $\Delta_{\alpha}(a) \in {[0,1]}^{Q \times Q}$ with
$\Delta_{\alpha}(a)[q,q'] = \sum_{\m \in \moves(q)} \alpha(q)(\m) \cdot \m(a, q')$.
Viewing subdistributions~$\mu$ as row vectors, we have:
\begin{equation} \label{eq-Succ-as-prod}
\Succ(\mu, \alpha, a) = \mu \cdot \Delta_{\alpha}(a)
\end{equation}

For a trace-based strategy $\alpha : \Label^{*} \times Q \to \dists(\moves)$ and a trace $w \in \Label^*$,
define the local strategy $\alpha[w] : Q \to \dists(\moves)$ with $\alpha[w](q) = \alpha(w,q)$ for all $q \in Q$.
We have the following lemma.
\begin{lem}%
\label{lem-subDis-as-matrix-prod}
Let $\D = \tuple{Q,\mu_0,\Label,\delta}$ be an MDP\@.
Let $\alpha : \Label^{*} \times Q \to \dists(\moves)$ be a trace-based strategy.
Let $w \in \Label^*$ and $a \in \Label$.
Then:
\[ \subdis_{\D,\alpha}(w a) = \subdis_{\D,\alpha}(w) \cdot \Delta_{\alpha[w]}(a)
\]
\end{lem}
\begin{proof}
Let $q' \in Q$.
We have:
\begin{align*}
& \subdis_{\D,\alpha}(w a)(q') \\
& = \sum_{\rho \in \Path(w)} \Prob_{\D,\alpha}(\rho a q')
 && \text{definition of $\subdis$} \\
& = \sum_{q \in Q} \sum_{\rho \in \Path(w,q)} \Prob_{\D,\alpha}(\rho) \cdot
     \sum_{\m \in \moves(q)} \alpha(\rho)(\m) \cdot \m(a,q')
 && \text{definition of $\Prob$} \\
& = \sum_{q \in Q} \sum_{\rho \in \Path(w,q)} \Prob_{\D,\alpha}(\rho) \cdot
     \sum_{\m \in \moves(q)} \alpha(w,q)(\m) \cdot \m(a,q')
 && \text{$\alpha$ is trace-based} \\
& = \sum_{q \in Q} \subdis_{\D,\alpha}(w)(q) \cdot
     \sum_{\m \in \moves(q)} \alpha(w,q)(\m) \cdot \m(a,q')
 && \text{definition of $\subdis$} \\
& = \sum_{q \in Q} \subdis_{\D,\alpha}(w)(q) \cdot
     \sum_{\m \in \moves(q)} \alpha[w](q)(\m) \cdot \m(a,q')
 && \text{definition of $\alpha[w]$} \\
& = \sum_{q \in Q} \subdis_{\D,\alpha}(w)(q) \cdot
     \Delta_{\alpha[w]}(a)[q,q']
 && \text{definition of $\Delta_{\alpha[w]}(a)$} \\
& = \left( \subdis_{\D,\alpha}(w) \cdot \Delta_{\alpha[w]}(a) \right) (q')
\tag*{\qedhere}
\end{align*}
\end{proof}

The following lemma is based on the idea that, using Lemma~\ref{lem-subDis-as-matrix-prod}, we can ``slice'' a strategy into local strategies, and conversely we can compose local strategies to a strategy.
\begin{lem}%
\label{lem-link-trace-local}
Let $\D = \tuple{Q,\mu_0,\Label,\delta}$ be an MDP\@.
Let $w = a_1 a_2 \cdots a_n \in \Label^*$.
Let $\mu_1, \mu_2, \ldots, \mu_n$ be subdistributions over~$Q$.
Then there is a strategy $\alpha : \Path(\D) \to \dists(\moves)$ with
\[
 \mu_i = \subdis_{\D,\alpha}(a_1 a_2 \cdots a_i) \qquad \text{for all $i \in \{0, 1, \ldots, n\}$}
\]
if and only if there are local strategies $\alpha_0, \alpha_1, \ldots, \alpha_{n-1} : Q \to \dists(\moves)$ with
\[
 \mu_{i+1} = \Succ(\mu_i, \alpha_i, a_{i+1}) \qquad \text{for all $i \in \{0, 1, \ldots, n-1\}$}.
\]
\end{lem}

\begin{proof}
We prove the two implications from the lemma in turn.
\begin{itemize}[align=left]
\item[``$\Longrightarrow$'':]
Let $\alpha$ be a strategy with
$
 \mu_i = \subdis_{\D,\alpha}(a_1 a_2 \cdots a_i)
$
for all $i \in \{0, 1, \ldots, n\}$.
By Lemma~\ref{lem-trace-based-enough} we can assume that $\alpha$ is trace-based.
For all $i \in \{0, 1, \ldots, n-1\}$ define a local strategy $\alpha_i$ with $\alpha_i = \alpha[a_1 a_2 \cdots a_i]$.
Then we have for all $i \in \{0, 1, \ldots, n-1\}$:
\begin{align*}
\mu_{i+1}
& = \subdis_{\D,\alpha}(a_1 a_2 \cdots a_{i+1}) && \text{definition of $\alpha$} \\
& = \subdis_{\D,\alpha}(a_1 a_2 \cdots a_{i}) \cdot \Delta_{\alpha[a_1 a_2 \cdots a_i]}(a_{i+1})
    && \text{Lemma~\ref{lem-subDis-as-matrix-prod}} \\
& = \mu_i \cdot \Delta_{\alpha_i}(a_{i+1})
    && \text{definitions of $\alpha, \alpha_i$} \\
& = \Succ(\mu_i, \alpha_i, a_{i+1})
    && \text{by~\eqref{eq-Succ-as-prod}}
\end{align*}
\item[``$\Longleftarrow$'':]
Let $\alpha_0, \alpha_1, \ldots, \alpha_{n-1}$ be local strategies with
$
 \mu_{i+1} = \Succ(\mu_i, \alpha_i, a_{i+1})
$
for all $i \in \{0, 1, \ldots, n-1\}$.
Define a trace-based strategy~$\alpha$ 
such that
$\alpha[a_1 a_2 \cdots a_i] = \alpha_i$ for all $i \in \{0, 1, \ldots, n-1\}$.
(This condition need not completely determine~$\alpha$.)
We prove by induction on~$i$ that
$
 \mu_i = \subdis_{\D,\alpha}(a_1 a_2 \cdots a_i)
$
for all $i \in \{0, 1, \ldots, n\}$.
For $i=0$ this is trivial.
For the step, we have:
\begin{align*}
\mu_{i+1}
& = \Succ(\mu_i, \alpha_i, a_{i+1}) && \text{definition of $\alpha_i$} \\
& = \mu_i \cdot \Delta_{\alpha_i}(a_{i+1}) && \text{by~\eqref{eq-Succ-as-prod}} \\
& = \subdis_{\D,\alpha}(a_1 a_2 \cdots a_i) \cdot \Delta_{\alpha_i}(a_{i+1}) && \text{induction hypothesis} \\
& = \subdis_{\D,\alpha}(a_1 a_2 \cdots a_i) \cdot \Delta_{\alpha[a_1 a_2 \cdots a_i]}(a_{i+1}) && \text{definition of $\alpha$} \\
& = \subdis_{\D,\alpha}(a_1 a_2 \cdots a_{i+1}) && \text{Lemma~\ref{lem-subDis-as-matrix-prod}}\qedhere
\end{align*}
\end{itemize}
\end{proof}

\noindent
Let   $\D = \tuple{Q_\D,\mu^{\D}_{0},\Label,\deltaD}$ and $\E = \tuple{Q_\E,\mu^{\E}_{0},\Label,\deltaE}$ be two MDPs over
the same set~$\Label$ of labels.
A \emph{bisimulation} is a relation~$\R \subseteq \subdists(Q_\D) \times \subdists(Q_\E)$ such that whenever
 $\mu_\D \mathrel{\R} \mu_\E$ then
\begin{itemize}
\item $\norm{\mu_\D} = \norm{\mu_\E}$;
\item for all local strategies~$\alpha_\D$ there exists a local strategy~$\alpha_\E$ such that for all $a \in \Label$ we have $\Succ(\mu_\D, \alpha_\D, a) \mathrel{\R} \Succ(\mu_\E, \alpha_\E, a)$;
\item for all local strategies~$\alpha_\E$ there exists a local strategy~$\alpha_\D$ such that for all $a \in \Label$ we have $\Succ(\mu_\D, \alpha_\D, a) \mathrel{\R} \Succ(\mu_\E, \alpha_\E, a)$.
\end{itemize}
As usual, a union of bisimulations is a bisimulation.
Denote by~$\mathord{\sim}$ the union of all bisimulations, i.e., $\mathord{\sim}$ is the largest bisimulation.
We write $\D \sim \E$ if $\mu^\D_{0} \sim \mu^{\E}_{0}$.
In general, the set~$\sim$ is uncountably infinite, so methods for computing state-based bisimulation (e.g., partition refinement) are not applicable.

Proposition~\ref{prop-link-trace-refinement-bisim} below establishes a link between trace refinement and bisimulation.
An intuitive interpretation of the proposition is that if $\D$ is an MDP and $\C$ an MC, then the best way of disproving bisimilarity between $\D$ and~$\C$ is to exhibit a sequence of local strategies in~$\D$ so that the resulting behaviour of~$\D$ cannot be matched by~$\C$.
Using Lemma~\ref{lem-link-trace-local} this sequence of local strategies can be assembled to a strategy for~$\D$, which then witnesses that $\D \not\sqsubseteq \C$.

\begin{prop}%
\label{prop-link-trace-refinement-bisim}
Let $\D$ be an MDP and $\C$ be an MC\@.
Then $\D \sim \C$ if and only if $\D \sqsubseteq \C$.
\end{prop}

\begin{proof}
Let $\D = \tuple{Q_\D,\mu^{\D}_{0},\Label,\delta^{\D}}$ and $\C = \tuple{Q_\C,\mu^{\C}_{0},\Label,\delta^{\C}}$.
\begin{itemize}[align=left]
\item[``$\Longrightarrow$'':]
Let $\D \sim \C$.
Hence $\mu^{\D}_{0} \sim \mu^{\C}_{0}$.
We show that $\D \sqsubseteq \C$.
Let $\alpha^{\D}$ be a strategy for~$\D$.
Let $w = a_1 a_2 \cdots a_n \in \Label^*$.
Let $\mu^{\D}_{0}, \mu^{\D}_{1}, \ldots, \mu^{\D}_{n}$ be the subdistributions with
 $\mu^{\D}_{i} = \subdis_{\D,\alpha^{\D}}(a_1 a_2 \cdots a_i)$ for all~$i$.
By Lemma~\ref{lem-link-trace-local} there exist local strategies $\alpha^{\D}_0, \alpha^{\D}_1, \ldots, \alpha^{\D}_{n-1}$ with
$\mu^{\D}_{i+1} = \Succ(\mu^{\D}_{i}, \alpha^{\D}_i, a_{i+1})$ for all~$i$.
Since $\mu^{\D}_{0} \sim \mu^{\C}_{0}$,
there exist local strategies $\alpha^{\C}_0, \alpha^{\C}_1, \ldots, \alpha^{\C}_{n-1}$ for~$\C$ and subdistributions $\mu^{\C}_1, \mu^{\C}_2, \ldots, \mu^{\C}_n$ with
$\mu^{\C}_{i+1} = \Succ(\mu^{\C}_{i}, \alpha^{\C}_i, a_{i+1})$ for all~$i$ and
$\mu^{\D}_i \sim \mu^{\C}_i$ for all~$i$.
Since $\C$ is an MC, the local strategies~$\alpha^{\C}_i$ are, in fact, irrelevant.
By Lemma~\ref{lem-link-trace-local} we have $\mu^{\C}_i = \subdis_{\C}(a_1 a_2 \cdots a_i)$ for all~$i$.
So we have:
\begin{align*}
\Tr_{\D,\alpha^{\D}}(w)
& = \norm{\subdis_{\D,\alpha^{\D}}(w)} && \text{by~\eqref{eq-subdis=trace-probability}} \\
& = \norm{\mu^{\D}_n} && \mu^{\D}_n = \subdis_{\D,\alpha^{\D}}(w) \\
& = \norm{\mu^{\C}_n} && \mu^{\D}_n \sim \mu^{\C}_n \\
& = \norm{\subdis_{\C}(w)} && \mu^{\C}_n = \subdis_{\C}(w) \\
& = \Tr_{\C}(w) && \text{by~\eqref{eq-subdis=trace-probability}}
\end{align*}
Since $\alpha^{\D}$ and~$w$ were chosen arbitrarily, we conclude that $\D \sqsubseteq \C$.
\item[``$\Longleftarrow$'':]
Let $\D \sqsubseteq \C$.
We show $\mu^{\D}_{0} \sim \mu^{\C}_{0}$.
Define a relation $\R \subseteq \subdists(Q_\D) \times \subdists(Q_\C)$
such that $\mu^{\D} \mathrel{\R} \mu^{\C}$ if and only if there exist a strategy~$\alpha^{\D}$ for~$\D$ and a trace~$w$ with
$\mu^{\D} = \subdis_{\D,\alpha^{\D}}(w)$ and $\mu^{\C} = \subdis_{\C}(w)$.
We claim that $\R$ is a bisimulation.
To prove the claim, consider any $\mu^{\D}, \mu^{\C}$ with $\mu^{\D} \mathrel{\R} \mu^{\C}$.
Then there exist a strategy~$\alpha^{\D}$ for~$\D$ and a trace~$w$ with
$\mu^{\D} = \subdis_{\D,\alpha^{\D}}(w)$ and $\mu^\C = \subdis_{\C}(w)$.
Since $\D \sqsubseteq \C$, we have $\Tr_{\D,\alpha^{\D}}(w) = \Tr_{\C}(w)$.
So we have:
\begin{align*}
\norm{\mu^{\D}}
& = \norm{\subdis_{\D,\alpha^{\D}}(w)} && \mu^{\D} = \subdis_{\D,\alpha^{\D}}(w) \\
& = \Tr_{\D,\alpha^{\D}}(w) && \text{by~\eqref{eq-subdis=trace-probability}} \\
& = \Tr_{\C}(w)       && \text{as argued above} \\
& = \norm{\subdis_{\C}(w)} && \text{by~\eqref{eq-subdis=trace-probability}} \\
& = \norm{\mu^{\C}} && \mu^{\C} = \subdis_{\C}(w)
\end{align*}
This proves the first condition for $\R$ being a bisimulation.

For the rest of the proof assume $w = a_1 a_2 \cdots a_n$.
Write $\mu^{\D}_n = \mu^{\D}$ and $\mu^{\C}_n = \mu^{\C}$.
Let $\alpha^{\D}_n$ be a local strategy for~$\D$.
Let $a_{n+1} \in \Label$.
Define $\mu^{\D}_{n+1} = \Succ(\mu^{\D}_n, \alpha^{\D}_n, a_{n+1})$,
   and $\mu^{\C}_{n+1} = \Succ(\mu^{\C}_n, \alpha^{\C}_n, a_{n+1})$ for an arbitrary (and unimportant as $\C$ is an MC) local strategy $\alpha^{\C}_n$ for~$\C$.
For the second and the third condition of $\R$ being a bisimulation we need to prove
$\mu^{\D}_{n+1} \mathrel{\R} \mu^{\C}_{n+1}$.
Define $\mu^{\D}_1, \mu^{\D}_2, \ldots, \mu^{\D}_{n-1}$ such that
$\mu^{\D}_i = \subdis_{\D,\alpha^{\D}}(a_1 a_2 \cdots a_i)$ for all $i \in \{0, 1, \ldots, n\}$.
By Lemma~\ref{lem-link-trace-local} there are local strategies $\alpha^{\D}_0, \alpha^{\D}_1, \ldots, \alpha^{\D}_{n-1}$
such that $\mu^{\D}_{i+1} = \Succ(\mu^{\D}_i, \alpha^{\D}_i, a_{i+1})$ for all $i \in \{0, 1, \ldots, n-1\}$.
We also have $\mu^{\D}_{n+1} = \Succ(\mu^{\D}_n, \alpha^{\D}_n, a_{n+1})$,
so again by Lemma~\ref{lem-link-trace-local} there is a strategy~$\beta^{\D}$ with
$\mu^{\D}_i = \subdis_{\D,\beta^{\D}}(a_1 a_2 \cdots a_i)$ for all $i \in \{0, 1, \ldots, n+1\}$. 
In particular, $\mu^{\D}_{n+1} = \subdis_{\D,\beta^{\D}}(w a_{n+1})$.
Similarly, we have $\mu^{\C}_{n+1} = \subdis_{\C}(w a_{n+1})$.
Thus, $\mu^{\D}_{n+1} \mathrel{\R} \mu^{\C}_{n+1}$.
Hence we have proved that $\R$ is a bisimulation.

Considering the empty trace, we see that $\mu^{\D}_0 \mathrel{\R} \mu^{\C}_0$.
Since $\R \subseteq \mathord{\sim}$, we also have $\mu^{\D}_0 \sim \mu^{\C}_0$, as desired. \qedhere
\end{itemize}
\end{proof}

\subsection{A Necessary and Sufficient Condition for Bisimilarity}%
\label{subsec:condition_bisimilar}

In the following we consider MDPs $\D = \tuple{Q,\mu^{\D}_{0},\Label,\delta}$ and $\E = \tuple{Q,\mu^{\E}_{0},\Label,\delta}$ over the same state space.
This is without loss of generality, since we might take the disjoint union of the state spaces.
Since $\D$ and~$\E$ differ only in the initial distribution, we will focus on~$\D$.

Let $B \in \Reals^{Q \times k}$ with $k \ge 1$.
Assume the label set is $\Label = \{a_1, \ldots, a_{|\Label|}\}$.
For $\mu \in \subdists(Q)$ and a local strategy~$\alpha$ 
we define a point $p(\mu, \alpha) \in \Reals^{{|\Label|} \cdot k}$ such that
\[
p(\mu, \alpha) \ = \ \big( \mu \Delta_{\alpha}(a_1) B \quad
                 \mu \Delta_{\alpha}(a_2) B \quad
                 \cdots \quad
                 \mu \Delta_{\alpha}(a_{|\Label|}) B \big).
\]
For the reader's intuition, we remark that we will choose matrices $B \in \Reals^{Q \times k}$ so that if two subdistributions $\mu_\D, \mu_\E$ are bisimilar then $\mu_\D B = \mu_\E B$.
(In fact, one can compute~$B$ so that the converse holds as well, i.e., $\mu_\D \sim \mu_\E$ if and only if $\mu_\D B = \mu_\E B$.)
It follows that, for subdistributions $\mu_\D, \mu_\E$ and local strategies $\alpha_\D, \alpha_\E$,
if $\Succ(\mu_\D, \alpha_\D, a) \sim \Succ(\mu_\E, \alpha_\E, a)$ holds for all $a \in \Label$ then $p(\mu_\D,\alpha_\D) = p(\mu_\E,\alpha_\E)$.
Let us also remark that for fixed $\mu \in \subdists(Q)$, the set $P_\mu = \{p(\mu,\alpha) \mid \alpha \text{ is a local strategy}\} \subseteq \Reals^{{|\Label|} \cdot k}$ is a (bounded and convex) polytope.
As a consequence, if $\mu_\D \sim \mu_\E$ then the polytopes $P_{\mu_\D}$ and~$P_{\mu_\E}$ must be equal.
In the next paragraph we define \emph{``extremal''} strategies~$\halpha$, which intuitively are local strategies such that $p(\mu,\halpha)$ is a vertex of the polytope~$P_{\mu}$.

Let $\vec{v} \in \Reals^{{|\Label|} \cdot k}$ be a \emph{column} vector;
we denote column vectors in boldface.
We view~$\vec{v}$ as a ``direction''.
Recall that~$d_q$ is the Dirac distribution on the state~$q$.
A pure local strategy~$\halpha$ is \emph{extremal in direction~$\vec{v}$ with respect to~$B$} if
\begin{align}
 & p(d_q, \alpha) \vec{v} \le p(d_q, \halpha) \vec{v}  \label{eq-remove-v1} \\
 & p(d_q, \alpha) \vec{v}  =  p(d_q, \halpha) \vec{v} \text{\quad implies\quad}
   p(d_q, \alpha)          =  p(d_q, \halpha) \label{eq-remove-v2}
\end{align}
for all states~$q \in Q$ and all pure local strategies~$\alpha$.

By linearity, if~\eqref{eq-remove-v1} and~\eqref{eq-remove-v2} hold for all pure local strategies~$\alpha$ then~\eqref{eq-remove-v1} and~\eqref{eq-remove-v2} hold for all local strategies~$\alpha$.
We say a local strategy $\halpha$ is \emph{extremal with respect to~$B$} if there is a direction~$\vec{v}$ such that $\halpha$ is extremal in direction~$\vec{v}$ with respect to~$B$.

In the following we prove some facts about extremal local strategies that will be needed later.

\begin{lem}%
\label{lem-remove-v}
Let $\D = \tuple{Q,\mu_{0},\Label,\delta}$ be an MDP\@.
Let $B \in \Reals^{Q \times k}$ with $k \ge 1$.
Let $\mu \in \subdists(Q)$.
Let $\alpha, \halpha$ be local strategies.
Suppose $\vec{v} \in \Reals^{{|\Label|} \cdot k}$ is a direction in which $\halpha$ is extremal and
$p(\mu, \alpha) \vec{v} = p(\mu, \halpha) \vec{v}$.
Then $p(\mu, \alpha) = p(\mu, \halpha)$.
\end{lem}
\begin{proof}
We have:
\begin{align*}
\sum_{q \in Q} \mu(q) \cdot p(d_q, \alpha) \vec{v}
& = p(\mu, \alpha) \vec{v} && \text{definition of~$p$} \\
& = p(\mu, \halpha) \vec{v} && \text{assumption on~$\halpha$} \\
& = \sum_{q \in Q} \mu(q) \cdot p(d_q, \halpha) \vec{v} && \text{definition of~$p$}
\end{align*}
With~\eqref{eq-remove-v1} it follows that for all $q \in \Supp(\mu)$ we have $p(d_q, \alpha) \vec{v} = p(d_q, \halpha) \vec{v}$.
Hence by~\eqref{eq-remove-v2} we obtain $p(d_q, \alpha) = p(d_q, \halpha)$ for all $q \in \Supp(\mu)$.
Thus:
\[
p(\mu, \alpha)
= \sum_{q \in Q} \mu(q) \cdot p(d_q, \alpha)
= \sum_{q \in Q} \mu(q) \cdot p(d_q, \halpha)
= p(\mu, \halpha)
\qedhere
\]
\end{proof}

For a subdistribution~$\mu$ define the bounded, convex polytope $P_{\mu} \subseteq \Reals^{{|\Label|} \cdot k}$ with
\[
P_{\mu} = \{p(\mu, \alpha) \mid \alpha : Q \to \dists(\moves)\}.
\]
Comparing two polytopes $P_{\mu_\D}$ and~$P_{\mu_\E}$ for subdistributions $\mu_\D, \mu_\E$ will play a key role for deciding bisimulation.
First we prove the following lemma, which states that any vertex of the polytope~$P_\mu$ can be obtained by applying an extremal local strategy.
Although this is intuitive, the proof is not very easy.

\begin{lem}%
\label{lem-vertex-extremal}
Let $\D = \tuple{Q,\mu_{0},\Label,\delta}$ be an MDP\@.
Let $B \in \Reals^{Q \times k}$ with $k \ge 1$.
Let $\mu \in \subdists(Q)$.
If $x \in P_{\mu}$ is a vertex of~$P_{\mu}$ then there is an extremal local strategy~$\halpha$ with $x = p(\mu, \halpha)$.
\end{lem}
\begin{proof}
Let $x \in P_{\mu}$ be a vertex of~$P_{\mu}$.
Let $\alpha_1 : Q \to \dists(\moves)$ be a local strategy so that $x = p(\mu, \alpha_1)$.
Since $x$ is a vertex, we can assume that $\alpha_1$ is pure.
Since $x$ is a vertex of~$P_{\mu}$, there is a hyperplane $H \subseteq \Reals^{{|\Label|} \cdot k}$ such that $\{x\} = P_{\mu} \cap H$.
Let $\vec{v}_1 \in \Reals^{{|\Label|} \cdot k}$ be a normal vector of~$H$.
Since $\{x\} = P_{\mu} \cap H$, we have $x \vec{v}_1 = \max_{y \in P_{\mu}} y \vec{v}_1$ or $x \vec{v}_1 = \min_{y \in P_{\mu}} y \vec{v}_1$; without loss of generality, say $x \vec{v}_1 = \max_{y \in P_{\mu}} y \vec{v}_1$.
Since $\{x\} = P_{\mu} \cap H$, we have for all $q \in \Supp(\mu)$ and all $\alpha$:
\begin{equation} \label{eq-lem-vertex-extremal-suppD}
p(d_q, \alpha) \vec{v}_1 = p(d_q, \alpha_1) \vec{v}_1 \quad \text{implies} \quad p(d_q, \alpha) = p(d_q, \alpha_1).
\end{equation}

For all $q \in Q \setminus \Supp(\mu)$, redefine the pure local strategy~$\alpha_1(q)$ so that all $q \in Q$ and all local strategies~$\alpha$ satisfy $p(d_q, \alpha) \vec{v}_1 \le p(d_q, \alpha_1) \vec{v}_1$.
Since $Q$~and~$\moves$ are finite, there is $\varepsilon > 0$ such that all $q \in Q$ and all \emph{pure} local strategies~$\alpha$  either satisfy $p(d_q, \alpha) \vec{v}_1 = p(d_q, \alpha_1) \vec{v}_1$ or $p(d_q, \alpha) \vec{v}_1 \le p(d_q, \alpha_1) \vec{v}_1 - \varepsilon$.

Define
\[
 \Sigma = \left\{ \alpha : Q \to \dists(\moves) \mid \forall\, q \in Q: p(d_q, \alpha) \vec{v}_1 = p(d_q, \alpha_1) \vec{v}_1 \right\}.
\]
Consider the bounded, convex polytope $P_2 \subseteq \Reals^{{|\Label|} \cdot k}$ defined by
\[
P_2 = \left\{
\sum_{q \in Q} p(d_q, \alpha) \; \middle\vert \;
\alpha \in \Sigma\right\} .
\]
By an argument similar to the one above, there are a pure local strategy $\halpha \in \Sigma$, a vertex $x_2 = \sum_{q \in Q} p(d_q, \halpha)$ of~$P_2$, and a vector $\vec{v}_2 \in \Reals^{{|\Label|} \cdot k}$ such that for all $q \in Q$ and all $\alpha \in \Sigma$, we have $p(d_q, \alpha) \vec{v}_2 \le p(d_q, \halpha) \vec{v}_2$, and if $p(d_q, \alpha) \vec{v}_2 = p(d_q, \halpha) \vec{v}_2$ then $p(d_q, \alpha) = p(d_q, \halpha)$.
By scaling down~$\vec{v}_2$ by a small positive scalar, we can assume that all $q \in Q$ and all local strategies~$\alpha$ satisfy
\begin{equation} \label{eq-small-epsilon}
 \left\lvert p(d_q, \alpha) \vec{v}_2 \right\rvert \le \frac{\varepsilon}{3}.
\end{equation}

Since $\halpha \in \Sigma$, all $q \in Q$ satisfy $p(d_q, \halpha) \vec{v}_1 = p(d_q, \alpha_1) \vec{v}_1$.
By~\eqref{eq-lem-vertex-extremal-suppD} all $q \in \Supp(\mu)$ satisfy $p(d_q, \halpha) = p(d_q, \alpha_1)$.
Hence:
\[
p(\mu, \halpha)
= \sum_{q \in \Supp(\mu)} \mu(q) p(d_q, \halpha)
= \sum_{q \in \Supp(\mu)} \mu(q) p(d_q, \alpha_1)
= p(\mu, \alpha_1) = x
\]

It remains to show that there is a direction~$\vec{v}$ in which $\halpha$ is extremal.
Take $\vec{v} = \vec{v}_1 + \vec{v}_2$.
Let $q \in Q$ and let $\alpha$ be a pure local strategy.
We consider two cases:
\begin{itemize}
\item
Assume $p(d_q, \alpha) \vec{v}_1 = p(d_q, \alpha_1) \vec{v}_1$.
Then there is $\beta \in \Sigma$ with $\alpha(q) = \beta(q)$, hence $p(d_q, \alpha) = p(d_q, \beta)$.
We have:
\begin{align*}
p(d_q, \alpha) \vec{v}
& = p(d_q, \beta) \vec{v} && p(d_q, \alpha) = p(d_q, \beta) \\
& = p(d_q, \beta) \vec{v}_1 + p(d_q, \beta) \vec{v}_2 && \text{definition of~$\vec{v}$} \\
& = p(d_q, \alpha_1) \vec{v}_1 + p(d_q, \beta) \vec{v}_2 && \beta \in \Sigma \\
& = p(d_q, \halpha) \vec{v}_1 + p(d_q, \beta) \vec{v}_2 && \halpha \in \Sigma \\
& \le p(d_q, \halpha) \vec{v}_1 + p(d_q, \halpha) \vec{v}_2 && \text{definition of $\halpha$} \\
& = p(d_q, \halpha) \vec{v} && \text{definition of~$\vec{v}$}
\end{align*}
Hence~\eqref{eq-remove-v1} holds for~$\halpha$.
To show~\eqref{eq-remove-v2}, assume $p(d_q, \alpha) \vec{v} = p(d_q, \halpha) \vec{v}$.
Then all terms in the computation above are equal, and $p(d_q, \beta) \vec{v}_2 = p(d_q, \halpha) \vec{v}_2$.
By the definition of~$\halpha$, this implies $p(d_q, \beta) = p(d_q, \halpha)$.
Hence $p(d_q, \alpha) = p(d_q, \beta) = p(d_q, \halpha)$.
Hence~\eqref{eq-remove-v2} holds for~$\halpha$.
\item
Assume $p(d_q, \alpha) \vec{v}_1 \ne p(d_q, \alpha_1) \vec{v}_1$.
By the definition of~$\varepsilon$ it follows $p(d_q, \alpha) \vec{v}_1 \le p(d_q, \alpha_1) \vec{v}_1 - \varepsilon$.
We have:
\begin{align*}
p(d_q, \alpha) \vec{v}
& = p(d_q, \alpha) \vec{v}_1 + p(d_q, \alpha) \vec{v}_2 && \text{definition of~$\vec{v}$} \\
& \le p(d_q, \alpha_1) \vec{v}_1 - \varepsilon + p(d_q, \alpha) \vec{v}_2 && \text{as argued above} \\
&  =  p(d_q, \halpha) \vec{v}_1 - \varepsilon + p(d_q, \alpha) \vec{v}_2 && \halpha \in \Sigma \\
& \le p(d_q, \halpha) \vec{v}_1 - \varepsilon + \frac{\varepsilon}{3} && \text{by~\eqref{eq-small-epsilon}} \\
& \le p(d_q, \halpha) \vec{v}_1 + p(d_q, \halpha) \vec{v}_2 - \varepsilon + \frac{\varepsilon}{3} + \frac{\varepsilon}{3} && \text{by~\eqref{eq-small-epsilon}} \\
& < p(d_q, \halpha) \vec{v}_1 + p(d_q, \halpha) \vec{v}_2 && \varepsilon > 0 \\
& = p(d_q, \halpha) \vec{v} && \text{definition of~$\vec{v}$}
\end{align*}
This implies~\eqref{eq-remove-v1} and~\eqref{eq-remove-v2} for~$\halpha$.
\end{itemize}
Hence, $\halpha$ is extremal in direction~$\vec{v}$.
\end{proof}

The following lemma states the intuitive fact that in order to compare the polytopes $P_{\mu_\D}$ and~$P_{\mu_\E}$, it suffices to compare the vertices obtained by applying extremal local strategies:

\begin{lem}%
\label{lem-check-extremal-enough}
Let $\D = \tuple{Q,\mu_{0},\Label,\delta}$ be an MDP\@.
Let $B \in \Reals^{Q \times k}$ with $k \ge 1$.
Then for all $\mu_\D, \mu_\E \in \subdists(Q)$ we have $P_{\mu_\D} = P_{\mu_\E}$ if and only if for all extremal local strategies~$\halpha$ we have $p(\mu_\D, \halpha) = p(\mu_\E, \halpha)$.
\end{lem}
\begin{proof}
We prove the two implications from the lemma in turn.
\begin{itemize}[align=left]
\item[``$\Longrightarrow$'':]
Suppose $P_{\mu_\D} = P_{\mu_\E}$.
Let $\halpha$ be a local strategy that is extremal in direction~$\vec{v}$.
Since $P_{\mu_\D} = P_{\mu_\E}$, there are $\alpha_{\E}$ and~$\alpha_{\D}$ such that
$p(\mu_\D, \halpha) = p(\mu_\E, \alpha_\E)$ and
$p(\mu_\E, \halpha) = p(\mu_\D, \alpha_\D)$.
We have:
\begin{align*}
p(\mu_\D, \halpha) \vec{v}
& = p(\mu_\E, \alpha_\E) \vec{v}
 && p(\mu_\D, \halpha) = p(\mu_\E, \alpha_\E) \\
& \le p(\mu_\E, \halpha) \vec{v}
 && \text{$\halpha$ is extremal in direction~$\vec{v}$} \\
& = p(\mu_\D, \alpha_\D) \vec{v}
 && \text{$p(\mu_\E, \halpha) = p(\mu_\D, \alpha_\D)$} \\
& \le p(\mu_\D, \halpha) \vec{v}
 && \text{$\halpha$ is extremal in direction~$\vec{v}$}
\end{align*}
So all inequalities are in fact equalities.
In particular, we have $p(\mu_\D, \halpha) \vec{v} = p(\mu_\D, \alpha_\D) \vec{v}$.
It follows:
\begin{align*}
p(\mu_\D, \halpha)
& = p(\mu_\D, \alpha_\D) && \text{Lemma~\ref{lem-remove-v}} \\
& = p(\mu_\E, \halpha)  && \text{definition of~$\alpha_\D$}
\end{align*}
\item[``$\Longleftarrow$'':]
Let $x$ be a vertex of~$P_{\mu_\D}$.
By Lemma~\ref{lem-vertex-extremal} there exists an extremal local strategy~$\halpha$ with $x = p(\mu_\D, \halpha)$.
By the assumption we have $p(\mu_\D, \halpha) = p(\mu_\E, \halpha)$.
Hence $x = p(\mu_\D, \halpha) = p(\mu_\E, \halpha) \in P_{\mu_\E}$.
Since $x$ is an arbitrary vertex of~$P_{\mu_\D}$, and $P_{\mu_\D}, P_{\mu_\E}$ are bounded, convex polytopes, it follows $P_{\mu_\D} \subseteq P_{\mu_\E}$.
The reverse inclusion is shown similarly. \qedhere
\end{itemize}
\end{proof}

\noindent
The following lemma shows that the alternation of quantifiers over local strategies can be replaced by quantifying over extremal local strategies:

\begin{lem}%
\label{lem-sim-pivot}
Let $\D = \tuple{Q,\mu_{0},\Label,\delta}$ be an MDP\@.
Let $B \in \Reals^{Q \times k}$ with $k \ge 1$.
Let $\mu_\D, \mu_\E \in \subdists(Q)$.
In the following let $\alpha_\D, \alpha_\E$ range over local strategies, $\halpha$~over extremal local strategies, and $a$ over~$\Label$.
Then
\begin{equation}
\begin{aligned}
 &\forall \alpha_\D \exists \alpha_\E \forall a : \mu_\D \Delta_{\alpha_\D}(a) B = \mu_\E \Delta_{\alpha_\E}(a) B \\
 \quad \land \quad &
 \forall \alpha_\E \exists \alpha_\D \forall a : \mu_\D \Delta_{\alpha_\D}(a) B = \mu_\E \Delta_{\alpha_\E}(a) B
\end{aligned}
\label{eq-lem-sim-pivot}
\end{equation}
holds if and only the following holds:
\[
\forall \halpha \forall a: \mu_\D \Delta_{\halpha}(a) B = \mu_\E \Delta_{\halpha}(a) B
\]
\end{lem}

\begin{proof}
We have:
\begin{align*}
& \forall \alpha_\D \exists \alpha_\E \forall a : \mu_\D \Delta_{\alpha_\D}(a) B = \mu_\E \Delta_{\alpha_\E}(a) B \\
\quad\Longleftrightarrow\quad &
\forall \alpha_\D \exists \alpha_\E : p(\mu_\D,\alpha_\D) = p(\mu_\E,\alpha_\E) && \text{definition of~$p$} \\
\quad\Longleftrightarrow\quad &
P_{\mu_\D} \subseteq P_{\mu_\E}
\intertext{
It follows:
}
\eqref{eq-lem-sim-pivot} 
\quad\Longleftrightarrow\quad &
P_{\mu_\D} = P_{\mu_\E} \\
\quad\Longleftrightarrow\quad &
\forall \halpha \forall a: \mu_\D \Delta_{\halpha}(a) B = \mu_\E \Delta_{\halpha}(a) B
&& \text{Lemma~\ref{lem-check-extremal-enough}}
\qedhere
\end{align*}

\end{proof}

The following proposition provides necessary and sufficient conditions for bisimilarity, which---as we will see---can be effectively checked.

\begin{prop}%
\label{prop-coNP-vector-space}
Let $\D = \tuple{Q,\mu_{0},\Label,\delta}$ be an MDP\@.
Let $B \in \Reals^{Q \times k}$ with $k \ge 1$.
\begin{enumerate}
\item[(1)] Suppose that for all $\mu_\D, \mu_\E \in \subdists(Q)$ with $\mu_\D \sim \mu_\E$ we have $\mu_\D B = \mu_\E B$.
    Then for all $\mu_\D, \mu_\E \in \subdists(Q)$ with $\mu_\D \sim \mu_\E$ we have $\mu_\D \Delta_{\halpha}(a) B = \mu_\E \Delta_{\halpha}(a) B$ for all extremal local strategies~$\halpha$ and all $a \in \Label$.
\item[(2)] Suppose that $B$ includes the column vector $\vec{1} = {(1 \ 1 \cdots 1)}^T$ (where the superscript~$T$ denotes transpose) and that for all extremal local strategies~$\halpha$ and all $a \in \Label$ the columns of $\Delta_{\halpha}(a) B$ are in the linear span of the columns of~$B$.
    Then for all $\mu_\D, \mu_\E \in \subdists(Q)$ with $\mu_\D B = \mu_\E B$ we have $\mu_\D \sim \mu_\E$.
\end{enumerate}
\end{prop}

\begin{proof}\leavevmode
\begin{itemize}[align=left]
\item[(1)]
Let $\mu_\D, \mu_\E \in \subdists(Q)$ with $\mu_\D \sim \mu_\E$.
By the definition of bisimulation and using~\eqref{eq-Succ-as-prod}, we obtain:
\begin{itemize}
\item for all local strategies~$\alpha_\D$ there exists a local strategy~$\alpha_\E$ such that for all $a \in \Label$ we have $\mu_\D \Delta_{\alpha_\D}(a) \sim \mu_\E \Delta_{\alpha_\E}(a)$;
\item for all local strategies~$\alpha_\E$ there exists a local strategy~$\alpha_\D$ such that for all $a \in \Label$ we have $\mu_\D \Delta_{\alpha_\D}(a) \sim \mu_\E \Delta_{\alpha_\E}(a)$.
\end{itemize}
Using our assumption on~$B$, we see that~\eqref{eq-lem-sim-pivot} from Lemma~\ref{lem-sim-pivot} holds for $\mu_\D, \mu_\E$.
By Lemma~\ref{lem-sim-pivot} we have $\mu_\D \Delta_{\halpha}(a) B = \mu_\E \Delta_{\halpha}(a) B$ for all extremal local strategies~$\halpha$ and all $a \in \Label$.
\item[(2)]
It suffices to show that the relation $\mathord{\sim_B} \subseteq \subdists(Q) \times \subdists(Q)$ defined by
\[
 \mu_\D \sim_B \mu_\E \quad \Longleftrightarrow \quad \mu_\D B = \mu_\E B
\]
is a bisimulation.
Let $\mu_\D \sim_B \mu_\E$, i.e., $\mu_\D B = \mu_\E B$.
Since $B$ includes the column~$\vec{1}$, we have $\norm{\mu_\D} = \norm{\mu_\E}$.
Since for all extremal local strategies~$\halpha$ and all $a \in \Label$ the columns of $\Delta_{\halpha}(a) B$ are in the linear span of the columns of~$B$, we have $\vec{0}^T = (\mu_\D - \mu_\E) B = (\mu_\D - \mu_\E) \Delta_{\halpha}(a) B$ for all extremal local strategies~$\halpha$ and all $a \in \Label$.
Lemma~\ref{lem-sim-pivot} implies that~\eqref{eq-lem-sim-pivot} holds for $\mu_\D, \mu_\E$.
Using~\eqref{eq-Succ-as-prod} and the definition of~$\mathord{\sim_B}$, we obtain:
\begin{itemize}
\item for all local strategies~$\alpha_\D$ there exists a local strategy~$\alpha_\E$ such that for all $a \in \Label$ we have $\Succ(\mu_\D, \alpha_\D, a) \sim_B \Succ(\mu_\E, \alpha_\E, a)$;
\item for all local strategies~$\alpha_\E$ there exists a local strategy~$\alpha_\D$ such that for all $a \in \Label$ we have $\Succ(\mu_\D, \alpha_\D, a) \sim_B \Succ(\mu_\E, \alpha_\E, a)$.
\end{itemize}
Thus the relation~$\mathord{\sim_B}$ is a bisimulation. \qedhere
\end{itemize}
\end{proof}

\subsection{A coNP Algorithm for Checking Bisimilarity of two MDPs}%
\label{subsec:algo_conp_bisimilarity}

Proposition~\ref{prop-coNP-vector-space}
suggests an algorithm for determining bisimilarity in a given MDP $\D = \tuple{Q,\mu_{0},\Label,\delta}$.
More concretely, we can compute a matrix~$B$ such that for all subdistributions $\mu_\D, \mu_\E$ we have $\mu_\D \sim \mu_\E$ if and only if $\mu_\D B = \mu_\E B$.
The algorithm initializes~$B$ with the column vector~$\vec{1}$ and henceforth maintains the invariant that $\mu_\D \sim \mu_\E$ implies $\mu_\D B = \mu_\E B$.
\begin{itemize}
\item
If there exists an extremal local strategy~$\halpha$ and a label $a \in \Label$ such that a column of $\Delta_{\halpha}(a) B$ is linearly independent of the columns of~$B$, then add this column to~$B$.
This maintains the invariant by Proposition~\ref{prop-coNP-vector-space}~(1).
Repeat.
\item
Otherwise (i.e., such $\halpha$ and~$a$ do not exist) terminate.
Then by Proposition~\ref{prop-coNP-vector-space}~(2) we have $\mu_\D B = \mu_\E B \ \Longrightarrow \ \mu_\D \sim \mu_\E$.
Together with the invariant we get $\mu_\D \sim \mu_\E \ \Longleftrightarrow \ \mu_\D B = \mu_\E B$, as claimed.
\end{itemize}
This algorithm terminates because $B$ can have at most $|Q|$ linearly independent columns.

Along these lines we can also prove the following theorem.

\begin{thm}%
\label{thm-coNP-result}
The problem that, given two MDPs~$\D$ and $\E$, asks whether $\D \sim \E$ is in {\sf coNP}.
\end{thm}

\begin{proof}
Without loss of generality we assume
$\D = \tuple{Q,\mu_{\D},\Label,\delta}$ and
$\E = \tuple{Q,\mu_{\E},\Label,\delta}$.
Hence we wish to decide in {\sf NP} whether $\mu_{\D} \not\sim \mu_{\E}$.

We proceed along the lines of the algorithm above.
Specifically, it follows from the arguments there that $\mu_\D \not\sim \mu_\E$ holds if and only if the following condition~$\Cond$ holds:
\begin{quote}
There are $k \in \{1, 2, \ldots, |Q|\}$ and $\vec{b}_0 = \vec{1}$ and $\vec{b}_1, \ldots, \vec{b}_{k-1} \in \Reals^Q$
and $i_0, i_1, \ldots, i_{k-1} \in \{0, 1, \ldots, k-2\}$
and $a_1, a_2, \ldots, a_{k-1} \in \Label$ and pure local strategies $\halpha_1, \halpha_2, \ldots, \halpha_{k-1}$ such that for all $j \in \{1, 2, \ldots, k-1\}$
\begin{itemize}
\item $\halpha_j$ is extremal with respect to the matrix formed by the column vectors $\vec{b}_0, \vec{b}_1, \ldots, \vec{b}_{j-1}$ and
\item $i_j < j$ and
\item $\vec{b}_j = \Delta_{\halpha_j}(a_j) \vec{b}_{i_j}$,
\end{itemize}
and $\mu_{\D} \vec{b}_{k-1} \ne \mu_{\E} \vec{b}_{k-1}$.
\end{quote}
It remains to argue that $\Cond$ can be checked in~{\sf NP}.
We can nondeterministically guess $k \le |Q|$ and $i_0, i_1, \ldots, i_{k-1} \le k-2$ and $a_1, a_2, \ldots, a_{k-1} \in \Label$ and pure local strategies $\halpha_1, \halpha_2, \ldots, \halpha_{k-1}$.
This determines $\vec{b}_1, \ldots, \vec{b}_{k-1}$.
All conditions in~$\Cond$ are straightforward to check in polynomial time, except the condition that for all $j \in \{1, 2, \ldots, k-1\}$ we have that $\halpha_j$ is extremal with respect to $\vec{b}_0, \vec{b}_1, \ldots, \vec{b}_{j-1}$.
In the remainder of the proof, we argue that this can also be checked in polynomial time.

Let $j \in \{1, 2, \ldots, k-1\}$.
Let $B \in \Reals^{Q \times j}$ be the matrix with columns $\vec{b}_0, \vec{b}_1, \ldots, \vec{b}_{j-1}$.
We want to check that $\halpha_j$ is extremal with respect to~$B$.
For all $q \in Q$, compute in polynomial time the set $\eqmoves(q) \subseteq \moves(q)$ defined by
\[
\eqmoves(q) = \{ \m \in \moves(q) \mid p(d_q, \alpha_{q,\m}) = p(d_q, \halpha_j) \},
\]
where $\alpha_{q,\m}$ is a pure local strategy with $\alpha_{q,\m}(q)(\m) = 1$ (it does not matter how $\alpha_{q,\m}(q')$ is defined for $q' \ne q$).
We want to verify that~\eqref{eq-remove-v1}~and~\eqref{eq-remove-v2} holds for~$\halpha_j$.
Hence we need to find~$\vec{v} \in \Reals^{|\Label| \cdot j}$ so that for all $q \in Q$ and all $\m \in \moves(q) \setminus \eqmoves(q)$ we have $p(d_q, \alpha_{q,\m}) \vec{v} < p(d_q, \halpha_j) \vec{v}$.
If such a vector~$\vec{v}$ exists, it can be scaled up by a large positive scalar so that we have:
\begin{equation} \label{eq-v-lp}
 p(d_q, \alpha_{q,\m}) \vec{v} + 1 \le p(d_q, \halpha_j) \vec{v} \quad \forall\, q \in Q \quad \forall\, \m \in \moves(q) \setminus \eqmoves(q)
\end{equation}
Hence it suffices to check if there exists a vector~$\vec{v}$ that satisfies~\eqref{eq-v-lp}.
This amounts to a feasibility check of a linear program of polynomial size.
Such a check can be carried out in polynomial time~\cite{Khachiyan79}.
\end{proof}

\subsection{An NC Algorithm for Trace Refinement}%
\label{subsec:algo_nc_trace_refinement}

In the following we consider an MDP $\D = \tuple{Q,\mu^{\D}_{0},\Label,\delta}$ and an MC $\C = \tuple{Q_\C,\mu^{\C}_{0},\Label,\delta_\C}$.
Without loss of generality, we assume $Q_\C \subseteq Q$.
Similarly, we also assume that $\delta_\C$ is a restriction of~$\delta$, and hence we write $\C = \tuple{Q_\C,\mu^{\C}_{0},\Label,\delta}$.
We may view subdistributions $\mu_\C \in \subdists(Q_\C)$ as $\mu_\C \in \subdists(Q)$ in the natural way.
The following proposition is analogous to Proposition~\ref{prop-coNP-vector-space}.
The key difference is that the need for considering \emph{extremal} strategies has disappeared.
This is due to the fact that only one of the two models is nondeterministic.

\begin{prop}%
\label{prop-coNP-vector-space-MC}
Let $\D = \tuple{Q,\mu^{\D}_{0},\Label,\delta}$ be an MDP
and $\C = \tuple{Q_\C,\mu^{\C}_{0},\Label,\delta}$ be an MC with $Q_\C \subseteq Q$.
Let $B \in \Reals^{Q \times k}$ with $k \ge 1$.
In the following let $\mu_\D$ range over~$\subdists(Q)$ and $\mu_\C$ over~$\subdists(Q_\C)$.
\begin{enumerate}
\item[(1)] Suppose that for all $\mu_\D, \mu_\C$ with $\mu_\D \sim \mu_\C$ we have $\mu_\D B = \mu_\C B$.
    Then for all $\mu_\D, \mu_\C$ with $\mu_\D \sim \mu_\C$ we have $\mu_\D \Delta_{\alpha}(a) B = \mu_\C \Delta_{\alpha}(a) B$ for all local strategies~$\alpha$ and all $a \in \Label$.
\item[(2)] Suppose that $B$ includes the column vector $\vec{1} = {(1 \ 1 \cdots 1)}^T$ 
    and that for all local strategies~$\alpha$ and all $a \in \Label$ the columns of $\Delta_{\alpha}(a) B$ are in the linear span of the columns of~$B$.
    Then for all $\mu_\D, \mu_\C$ with $\mu_\D B = \mu_\C B$ we have $\mu_\D \sim \mu_\C$.
\end{enumerate}
\end{prop}

\begin{proof}
The proof is similar to but simpler than the proof of Proposition~\ref{prop-coNP-vector-space}.
For completeness, we give it explicitly.
\begin{itemize}[align=left]
\item[(1)]
Let $\mu_\D \sim \mu_\C$.
By the definition of bisimulation and using~\eqref{eq-Succ-as-prod} we have $\mu_\D \Delta_{\alpha}(a) \sim \mu_\C \Delta_{\alpha}(a)$ for all local strategies~$\alpha$ and all $a \in \Label$.
By our assumption on~$B$, we have $\mu_\D \Delta_{\alpha}(a) B = \mu_\C \Delta_{\alpha}(a) B$ for all local strategies~$\alpha$ and all $a \in \Label$.
\item[(2)]
It suffices to show that the relation $\mathord{\sim_B} \subseteq \subdists(Q) \times \subdists(Q_\C)$ defined by
\[
 \mu_\D \sim_B \mu_\C \quad \Longleftrightarrow \quad \mu_\D B = \mu_\C B
\]
is a bisimulation.
Let $\mu_\D \sim_B \mu_\C$, i.e., $\mu_\D B = \mu_\C B$.
Since $B$ includes the column~$\vec{1}$, we have $\norm{\mu_\D} = \norm{\mu_\C}$.
Since for all local strategies~$\alpha$ and all $a \in \Label$ the columns of~$\Delta_{\alpha}(a) B$ are in the linear span of the columns of~$B$, we have $\vec{0}^T = (\mu_\D - \mu_\C) B = (\mu_\D - \mu_\C) \Delta_{\alpha}(a) B$ for all local strategies~$\alpha$ and all $a \in \Label$.
Using~\eqref{eq-Succ-as-prod} and the definition of~$\mathord{\sim_B}$, we see that
for all local strategies~$\alpha$ and all $a \in \Label$ we have $\Succ(\mu_\D, \alpha, a) \sim_B \Succ(\mu_\C, \alpha, a)$.
Thus the relation~$\mathord{\sim_B}$ is a bisimulation. \qedhere
\end{itemize}
\end{proof}

\begin{cor}%
\label{cor-coNP-vector-space-MC}
Let $\mathcal{V} \subseteq \Reals^Q$ be the smallest column-vector space with $\vec{1} \in \mathcal{V}$ and $\Delta_\alpha(a) \vec{u} \in \mathcal{V}$ for all $\vec{u} \in \mathcal{V}$, all labels $a \in \Label$, and all local strategies~$\alpha$.
Then for all $\mu_\D \in \subdists(Q)$ and all $\mu_\C \in \subdists(Q_\C)$ we have:
\[
\mu_\D \sim \mu_\C \quad \Longleftrightarrow\quad \mu_\D \vec{u} = \mu_\C \vec{u}\ \text{ for all } \vec{u} \in \mathcal{V}
\]
\end{cor}

Notice the differences to Proposition~\ref{prop-coNP-vector-space}: there we considered all extremal local strategies (potentially exponentially many), here we consider all local strategies (in general infinitely many).
However, we show that one can efficiently find few local strategies that span all local strategies.
This allows us to reduce (in logarithmic space) the bisimulation problem between an MDP and an MC to the bisimulation problem between two MCs,
which is equivalent to the trace-equivalence problem in MCs (by Proposition~\ref{prop-link-trace-refinement-bisim}).
The latter problem is known to be in~{\sf NC}~\cite{Tzeng96}.
Theorem~\ref{thm-MDP-MC} then follows with Proposition~\ref{prop-link-trace-refinement-bisim}.

\begin{thm}%
\label{thm-MDP-MC}
The problem~${\sf MDP\sqsubseteq MC}$ is in~{\sf NC}, hence in~{\sf P}.
\end{thm}

\begin{proof}
Let $\D = \tuple{Q,\mu^{\D}_{0},\Label,\delta}$ be an MDP
and $\C = \tuple{Q_\C,\mu^{\C}_{0},\Label,\delta}$ be an MC with $Q_\C \subseteq Q$.

Let $\alpha_0$ denote an arbitrary pure local strategy.
For each $q \in Q$ and each $\m \in \moves(q)$ denote by~$\alpha_{q,\m}$ the pure local strategy such that $\alpha_{q,\m}(q)(\m) = 1$ and $\alpha_{q,\m}(q') = \alpha_0(q')$ for all $q' \in Q \setminus \{q\}$.
Define
\begin{align*}
\Sigma &= \{ \alpha_0 \} \cup \{ \alpha_{q,\m} \mid q \in Q, \ \m \in \moves(q) \} && \text{and} \\
 \M &= \{ \Delta_{\alpha}(a) \in \Reals^{Q \times Q} \mid \alpha \in \Sigma,\ a \in \Label\}
 && \text{and} \\
 \M_\infty &= \left\{ \Delta_\alpha(a) \in \Reals^{Q \times Q} \;\middle\vert\; \alpha \text{ is a local strategy, $a \in \Label$}\right\}.
\end{align*}
The vector space $\V \subseteq \Reals^Q$ from Corollary~\ref{cor-coNP-vector-space-MC} is the smallest vector space with
\begin{itemize}
\item
$\vec{1} = {(1 \ 1 \cdots 1)}^T \in \V$ and
\item $M \vec{u} \in \V$, for all $\vec{u} \in \V$ and all $M \in \M_\infty$.
\end{itemize}

\noindent
We have $\M \subseteq \M_\infty$, where $|\M|$ is finite and $|\M_\infty|$ is infinite.
Every matrix in~$\M_\infty$ can be expressed as a linear combination of matrices from~$\M$:
Indeed, let $\alpha$ be a local strategy.
Then for all $a \in \Label$ we have:
\[
 \Delta_\alpha(a) = \Delta_{\alpha_0}(a)
  + \sum_{q \in Q}
  \left( - \Delta_{\alpha_0}(a) +
   \sum_{\m \in \moves(q)}
      \alpha(q)(\m) \cdot \Delta_{\alpha_{q,\m(q)}}(a)
  \right)
\]
So by linearity, the vector space~$\V$ is the smallest column-vector space such that
\begin{itemize}
\item
$\vec{1} = {(1 \ 1 \cdots 1)}^T \in \V$ and
\item $M \vec{u} \in \V$, for all $\vec{u} \in \V$ and all $M \in \M$.
\end{itemize}

\noindent
Define a finite set of labels $\Label' = \{b_{\alpha, a} \mid \alpha \in \Sigma, \ a \in \Label \}$, and for each $\alpha \in \Sigma$ and each $a \in \Label$ a matrix
\[
 \Delta'(b_{\alpha, a}) = \frac{1}{|\Sigma|} \Delta_\alpha(a).
\]
The matrix $\sum_{b \in \Label'} \Delta'(b)$ is stochastic.
Define the MCs $\D' = \tuple{Q,\mu^{\D}_{0},\Label',\delta'}$
and $\C' = \tuple{Q,\mu^{\C}_{0},\Label',\delta'}$
such that $\delta'$ induces the transition matrices $\Delta'(b)$ for all $b \in \Label'$.
The MCs $\D'$ and~$\C'$ are computable in logarithmic space.
Let $\V' \subseteq \Reals^{Q}$ be the smallest column-vector space such that
\begin{itemize}
\item
$\vec{1} = {(1 \ 1 \cdots 1)}^T \in \V$ and
\item $\Delta'(b) \vec{u} \in \V$, for all $\vec{u} \in \V$ and all $b \in \Label'$.
\end{itemize}
Since the matrices in~$\M$ and the matrices~$\Delta'(b)$ are scalar multiples of each other, we have $\V = \V'$.
It holds:
\begin{align*}
\D \sqsubseteq \C \quad
& \Longleftrightarrow\quad \D \sim \C \text{ in~$\D$}
 && \text{Proposition~\ref{prop-link-trace-refinement-bisim}} \\
& \Longleftrightarrow\quad \mu^\D_{0} \sim \mu^{\C}_{0} \text{ in~$\D$}
 && \text{definition} \\
& \Longleftrightarrow\quad \forall\, \vec{u} \in \V : \mu^{\D}_{0} \vec{u} = \mu^{\C}_{0} \vec{u}
 && \text{Corollary~\ref{cor-coNP-vector-space-MC}} \\
& \Longleftrightarrow\quad \forall\, \vec{u} \in \V' : \mu^{\D}_{0} \vec{u} = \mu^{\C}_{0} \vec{u}
 && \V = \V' \\
& \Longleftrightarrow\quad \mu^\D_{0} \sim \mu^{\C}_{0} \text{ in~$\D'$}
 && \text{Corollary~\ref{cor-coNP-vector-space-MC}} \\
& \Longleftrightarrow\quad \D' \sim \C' \text{ in~$\D'$}
 && \text{definition} \\
& \Longleftrightarrow\quad \D' \sqsubseteq \C'
 && \text{Proposition~\ref{prop-link-trace-refinement-bisim}}
\end{align*}
As mentioned in Section~\ref{sub-prel-trace-refine},
deciding whether $\D' \sqsubseteq \C'$ holds amounts to the trace-equivalence problem for MCs.
It follows from Tzeng~\cite{Tzeng96} that the latter is decidable in~{\sf NC}, hence in~{\sf P}.
\end{proof}


\section{Conclusions}%
\label{sec-conclusions}
We have settled the decidability and complexity status of most subproblems of trace refinement between two MDPs.
Key technical ingredients were links to a certain notion of bisimulation, linear-algebra arguments, and comparisons of polytopes.

As an open problem, we highlight the complexity of the distribution-based notion of bisimulation, which we have shown to be in {\sf coNP}.
Is this notion of bisimulation {\sf coNP}-complete or in~{\sf P}?

\subsubsection*{Acknowledgement.}
The authors thank anonymous referees for their helpful comments.

\bibliographystyle{plain}
\bibliography{references}

\end{document}